\documentclass[aps,prl,twocolumn,showpacs,psfig,%
superscriptaddress,longbibliography]{revtex4-1}

\usepackage{amsmath,amssymb}
\usepackage{braket}
\usepackage{bm}
\usepackage{color}
\usepackage{etoolbox}
\usepackage{epstopdf}
\usepackage{float}
\usepackage{graphicx}
\usepackage[colorlinks=true,linkcolor=blue,citecolor=blue,urlcolor=blue]{hyperref}
\usepackage[latin9]{inputenc}
\usepackage{multirow}
\usepackage{soul}
\usepackage{subfigure}
\usepackage{times}
\usepackage{titlesec}
\usepackage{tikz}
\usepackage{txfonts}
\usepackage{xspace}
\usepackage{xfrac}

\newcommand{\updownarrows}{\mathbin\uparrow\hspace{-.5em}\downarrow}

\newcommand{\Eq}[1]{Eq.~\eqref{#1}}

\newcommand{\Fig}[1]{Fig.~\ref{#1}}

\graphicspath{{./Fig/}}

\begin{document}
\title{Quantum Many-Body Simulations of the 2D Fermi-Hubbard Model in Ultracold Optical Lattices}

\author{Bin-Bin Chen}
\affiliation{School of Physics and Key Laboratory of Micro-Nano Measurement-Manipulation and Physics (Ministry of Education), 
Beihang University, Beijing 100191, China}
\affiliation{Arnold Sommerfeld Center for Theoretical Physics, Center for NanoScience, and Munich Center for Quantum Science and Technology, Ludwig-Maximilians-Universit\"at M\"unchen, 80333 Munich, Germany}

\author{Chuang Chen}
\affiliation{Beijing National Laboratory for Condensed Matter Physics and Institute of Physics, Chinese Academy of Sciences, Beijing 100190, China}
\affiliation{School of Physical Sciences, University of Chinese Academy of Sciences, Beijing 100190, China}

\author{Ziyu Chen}
\affiliation{School of Physics and Key Laboratory of Micro-Nano Measurement-Manipulation and Physics (Ministry of Education), 
Beihang University, Beijing 100191, China}

\author{Jian Cui}
\affiliation{School of Physics and Key Laboratory of Micro-Nano Measurement-Manipulation and Physics (Ministry of Education), 
Beihang University, Beijing 100191, China}

\author{Yueyang Zhai}
\affiliation{Research Institute of Frontier Science, Beihang University, Beijing 100191, China}

\author{Andreas Weichselbaum}
\email{weichselbaum@bnl.gov}
\affiliation{Department of Condensed Matter Physics and Materials
Science, Brookhaven National Laboratory, Upton, New York 11973-5000, USA}
\affiliation{Arnold Sommerfeld Center for Theoretical Physics, Center for NanoScience, 
and Munich Center for Quantum Science and Technology, Ludwig-Maximilians-Universit\"at M\"unchen, 80333 Munich, Germany}

\author{Jan von Delft}
\affiliation{Arnold Sommerfeld Center for Theoretical Physics, Center for NanoScience, and Munich Center for Quantum Science and Technology, Ludwig-Maximilians-Universit\"at M\"unchen, 80333 Munich, Germany}

\author{Zi Yang Meng}
\email{zymeng@hku.hk}
\affiliation{Department of Physics and HKU-UCAS Joint Institute of Theoretical and Computational Physics, 
The University of Hong Kong, Pokfulam Road, Hong Kong SAR, China}
\affiliation{Beijing National Laboratory for Condensed Matter Physics and Institute of Physics, Chinese Academy of Sciences, Beijing 100190, China}
\affiliation{Songshan Lake Materials Laboratory, Dongguan, Guangdong 523808, China}

\author{Wei Li}
\email{w.li@buaa.edu.cn}
\affiliation{School of Physics and Key Laboratory of Micro-Nano Measurement-Manipulation and Physics (Ministry of Education), 
Beihang University, Beijing 100191, China}
\affiliation{International Research Institute of Multidisciplinary Science, Beihang University, Beijing 100191, China}

\begin{abstract}
Understanding quantum many-body states of correlated electrons is one main 
theme in modern condensed matter physics. Given that the Fermi-Hubbard model, 
the prototype of correlated electrons, has been recently realized in ultracold optical lattices, 
it is highly desirable to have controlled numerical methodology to provide precise  
finite-temperature results upon doping, to directly compare with experiments.
Here, we demonstrate the exponential tensor renormalization group (XTRG) 
algorithm [Phys. Rev. X \textbf{8}, 031082 (2018)], complemented 
with independent
determinant quantum Monte Carlo (DQMC) 
offer a powerful combination of tools
for this purpose. 
XTRG provides full and accurate access to the density matrix 
and thus various spin and charge correlations, 
down to unprecedented low temperature 
of few percents of the fermion tunneling energy scale.
We observe excellent agreement with ultracold fermion measurements 
at both half-filling and finite-doping, including the sign-reversal 
behavior in spin correlations due to formation of magnetic polarons, 
and the attractive hole-doublon and repulsive hole-hole pairs that are responsible for the 
peculiar bunching and antibunching behavior of the antimoments.
\end{abstract}
\date{\today}
\maketitle

\textit{Introduction.---} 
The Fermi-Hubbard model,
describing a paradigmatic quantum 
many-body system~\cite{Hubbard1963a,Gutzwiller1963a}, 
has relevance for a broad scope of correlation phenomena, 
ranging from high-temperature superconductivity~\cite{Lee2006}, 
metal-insulator transition~\cite{fazekas1999lecture}, 
quantum criticality~\cite{sachdev2011quantum}, to
interacting topological states of matter~\cite{wen2004quantum}.
Yet, puzzles remain in this strongly interacting many-body model 
after several decades of intensive investigations.
In solid-state materials, the Fermi-Hubbard model is 
often complicated by multi-band structures 
and interactions such as spin-orbital and Hund's couplings \cite{Georges2013}, etc. 
In this regard, recent progresses in two-dimensional (2D) fermionic optical lattices, 
where the interplay between the spin and charge degrees of freedom in the 
Fermi-Hubbard model has been implemented in a faithful way
\cite{Bakr2009,Parsons2015,Greif2016,Boll2016,Cheuk2016a,Cheuk2016b,Parsons2016}, 
enables a very clean and powerful platform for simulating its magnetic 
\cite{Greif2013,Hart2015,Mazurenko2017,Brown2017,Hilker2017,Chiu2019,
Salomon2019,Koepsell2019}
and transport properties \cite{Nichols2019,Brown2019}. 

\begin{figure}[tb]
\includegraphics[width=1\linewidth]{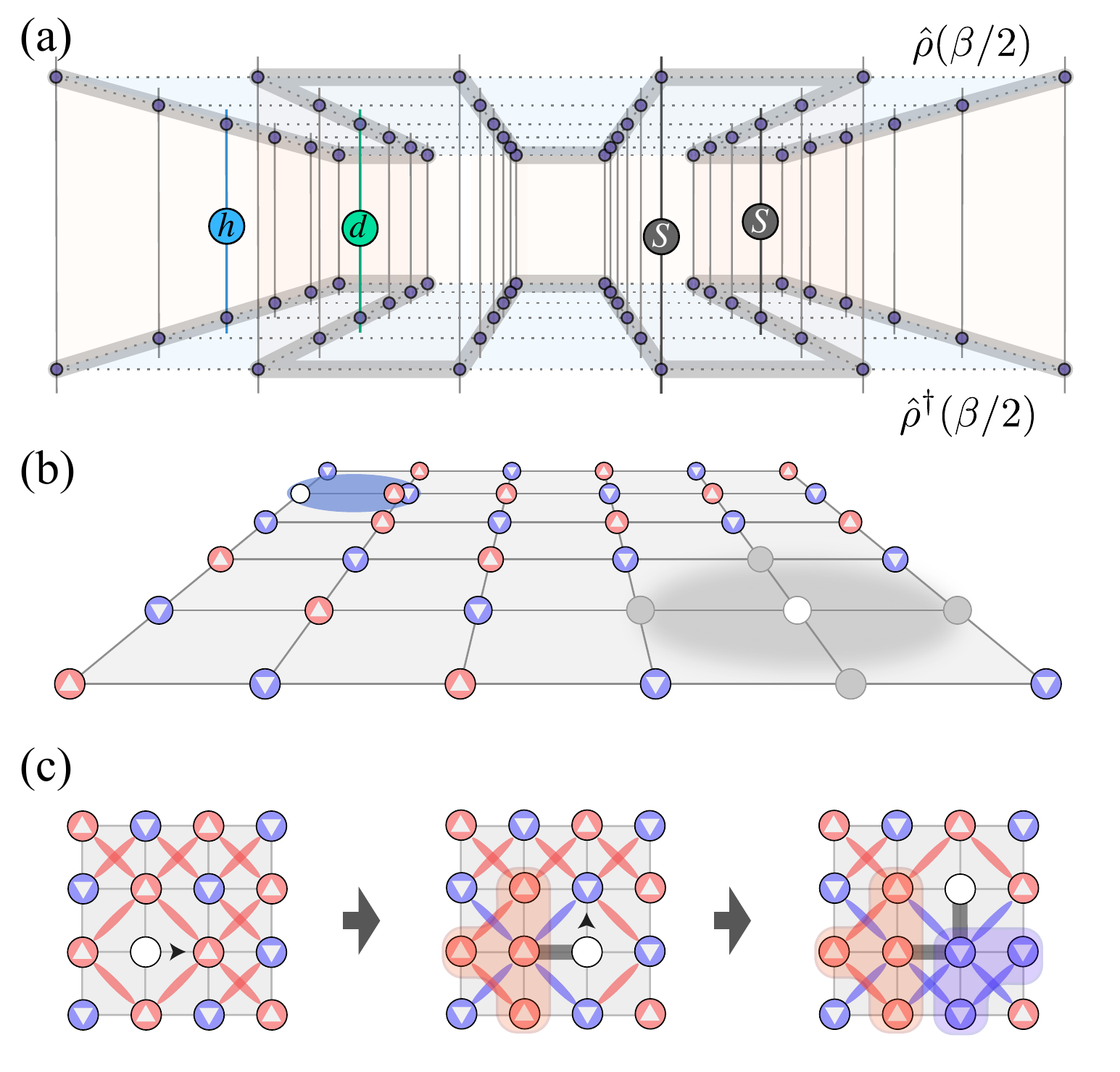}
\caption{(a) Bilayer calculation of the spin-spin $\langle \hat S_i\cdot \hat S_j\rangle$ and 
hole-doublon $\langle \hat h_i \cdot \hat d_j\rangle$ correlators by sandwiching corresponding 
operators in between 
$\hat \rho(\beta/2)$ and $\hat \rho^\dagger(\beta/2)$
where the snake-like ordering of sites for the XTRG
is indicated by thick gray line.
(b) In the low-temperature AF background
(blue down and red up spins), a magnetic polaron
(grey shaded region) 
emerges around a moving hole, where the spins around the hole 
can be in a superposition of spin-up and down states. The blue ellipse 
represents a hole-doublon pair showing a strong bunching effect. 
(c) A hole moves in the system along the path indicated by the grey string,
leading to a sign reversal of the diagonal spin correlation. The red- and blue-shaded regions 
illustrate the deformed magnetic background due to the interplay
between the hole and spins. Diagonal correlations
are indicated red (aligned) or blue (antialigned).
}
\label{Fig:XTRG}
\end{figure}

With the state-of-the-art quantum gas microscope techniques, 
single-site and spin-resolved imaging is now available, 
and ``snapshots" of correlated fermions have been studied
experimentally \cite{Bakr2009,Parsons2015,Greif2016,Cheuk2016a}. 
On top of that, detailed local 
spin and charge correlations 
\cite{Greif2013,Parsons2016,Cheuk2016b,Boll2016,Mazurenko2017,Koepsell2019},
as well as hidden orders revealed by pattern recognition \cite{Hilker2017,Chiu2019}, 
all inaccessible in traditional solid-state experiments, can be read out by the microscope.
As a highly controlled quantum simulator, ultracold
fermions in optical lattices therefore serve as a promising tool 
for resolving various intriguing theoretical proposals 
on the 2D Fermi-Hubbard model.
However, numerous challenges remain, both 
theoretically and experimentally. 
The currently lowest achievable temperature is f
$T/t \simeq 0.25$-0.5 (with $t$ the fermion tunneling energy)
on a finite-size system with about 70-80 $^6$Li atoms~\cite{Mazurenko2017,Chiu2019,Koepsell2019},
and $T/t\sim 1$ in $^{40}$K systems~\cite{Cheuk2016a,Hartke2020}.
These temperatures are still much higher than the estimated superconductivity 
transition temperature, $T_c/t \sim 0.05$, near the optimal 
doping of the square-lattice Hubbard model~\cite{Lee2006,KSChen2013}. 

On the theoretical side, it is then of vital importance to provide 
precise quantum many-body calculations in the 2D Hubbard model
for systems of similar size and fermion number as those studied experimentally.
Only with that, can one benchmark theory with the cold-atom experiment,
determine the effective temperature $T$ of the fermionic optical lattice system, 
explain experimental results, and provide accurate guidance for future progress. 
However, accurately computing properties of 2D Fermi-Hubbard model
at finite temperature and finite doping is difficult. 
Quantum Monte Carlo (QMC) methods suffer from the minus-sign problem, 
although with finite size and temperature, 
the QMC simulation can actually be performed, yielding 
unbiased results before one hits the ``exponential wall".  
In this regard, it is highly desirable to have an alternative and powerful method
whose accessible parameter space overlaps, on the one hand, with that of QMC
for benchmarking purposes, but which extends, on the other hand, to more difficult yet 
experimentally accessible regions.
In this letter, we demonstrate that the thermal tensor 
network approach stands out as the method of choice.

In fact, various tensor renormalization group (TRG) methods have been 
developed to compute the $T=0$ properties of the 2D Hubbard model 
\cite{Noack1994,Corboz2010,Kraus2010,Gu2010,LeBlanc2015,Zheng2017,Qin2019,Chung2020}.
However, the $T>0$ properties at finite doping are much less explored.
In this work, we generalize the exponential TRG (XTRG)
from spin system~\cite{Chen2018, Li2019} to strongly interacting fermions, 
and employ it to simulate the Fermi-Hubbard model at both half-filling and finite doping, 
down to a few percents of the tunneling energy $t$.
We compare the results obtained from both 
XTRG and determinant QMC (DQMC)~\cite{ChuangChen2019} in the parameter space 
where both methods are applicable, and find excellent
agreement between them as a consistency and sanity check.
Then we carry out XTRG$+$DQMC investigations of the 2D Hubbard model to cover the 
entire parameter space accessed by current cold-atom experiments. 
We find that the experimental quantum gas microscope data can be perfectly explained 
by our numerical simulations. 
The combined scheme of XTRG$+$DQMC therefore opens a route for systematic 
investigation of the finite-temperature phase diagram of the 2D Fermi-Hubbard model and 
constitutes an indispensable theoretical guide for ultracold fermion experiments.

\textit{The Fermi-Hubbard model.---} 
We consider interacting electrons on a 2D square lattice described by the Hamiltonian
\begin{equation}
\label{Eq:Hub}
H = -t\sum_{\langle {i},\hspace{.1em}{j} \rangle,\sigma} 
(\hat c_{{i},\sigma}^\dagger \hat c_{{j},\sigma}^{\phantom{\dagger}} + {h.c.}) 
+ U\sum_{i} \hat n_{i\uparrow} \hat n_{i\downarrow} 
- \mu\sum_{{i},\sigma}{ \hat n_{{i},\sigma}}, 
\end{equation}
with $t=1$ the nearest-neighbor hopping amplitude
(which thus sets the unit of energy, throughout), 
$U>0$ the on-site Coulomb repulsion, and $\mu$ the chemical potential controlling the electron filling.
The fermionic operator $\hat c_{{i},\sigma}$ annihilates an 
electron with spin $\sigma\in\{\uparrow,\downarrow\}$ on site ${i}$, 
and $\hat n_{{i},\sigma} \equiv \hat c_{{i},\sigma}^\dagger \hat c_{{i},\sigma}^{\phantom{\dagger}}$ 
is the local number operator.

In the large-$U$ limit $(U\gg t)$ and at half-filling ($\mu=U/2$), 
the Hubbard model can be effectively mapped to the Heisenberg model 
with interchange integral $J = 4t^2/U$, giving rise to a N\'eel-ordered 
ground state with 
strong antiferromagnetic (AF) correlations at low temperature 
[depicted schematically in \Fig{Fig:XTRG}(b)]. 
This has been demonstrated in many-body calculations~\cite{Varney2009} and 
recently observed in ultracold fermion experiments \cite{Mazurenko2017}.
In this work, we study the Fermi-Hubbard model with $U=7.2$,
a typical interaction strength used in recent experiments~
\cite{Cheuk2016a,Mazurenko2017,Chiu2019,Hartke2020},
and further tune the chemical potential  
$\mu<U/2$ to investigate the effect of hole doping.

\textit{Fermion XTRG.---}
Finite-temperature TRG methods have been proposed to 
compute the thermodynamics of interacting spins 
\cite{Verstraete2004,Zwolak2004,Feiguin2005,Li2011,Chen2017,Chen2018,Bruognolo2017,Chung2019}. 
However, the simulation of correlated fermions at finite temperature
has so far been either limited to relatively high temperature \cite{Khatami2011,Czarnik2014}
or to rather restricted geometries, like 1D chains \cite{Dong2017}.
XTRG employs a DMRG-like setup for both 1D and 2D systems \cite{Chen2018,Li2019} 
and cools down the systems exponentially fast in temperature. It has 
been shown to have great precision in simulating quantum spin systems 
on bipartite \cite{Chen2018} and frustrated lattices \cite{Chen2019,Li2020}.
It thus holds great promise to be generalized to correlated fermions.

As shown in \Fig{Fig:XTRG}(a), we represent the density matrix $\hat \rho(\beta/2)$ 
as a matrix product operator (MPO) defined on a 1D snake-like path [depicted as grey shaded lines
in \Fig{Fig:XTRG}(a)]. To accurately compute the expectation value of a observable $\hat O$, 
we adopt the bilayer technique \cite{Dong2017}, yielding 
$\langle\hat O\rangle = \frac{1}{\mathcal{Z}} \mathrm{Tr}[\hat\rho(\beta/2)\cdot\hat O\cdot\hat\rho^\dagger(\beta/2)], $
with $\mathcal{Z}=\mathrm{Tr}[\hat\rho(\beta/2)\cdot\hat\rho^\dagger(\beta/2)]$ 
the partition function. In practice, we adopt the QSpace framework \cite{Weichselbaum2012,Weichselbaum2020} 
and implement fermion and non-abelian
symmetries in our XTRG code (for technical details, see \cite{SM}).
We consider mainly two-site static correlators,
$\langle\hat O\rangle=\langle\hat O_{i} \cdot \hat O_{j} \rangle$, 
with 
$\hat O_{i}$ a local operator such as
the SU(2)~spinor $\hat S_{i} \equiv[\frac{-1}{\sqrt{2}} \hat 
c_{i\uparrow}^\dagger \hat c_{i\downarrow}^{\phantom{\dagger}}, 
\frac{1}{2} (\hat n_{i\uparrow} -\hat n_{i\downarrow}), 
\frac{1}{\sqrt{2}} \hat c_{i\downarrow}^\dagger 
\hat c_{i \uparrow}^{\phantom{\dagger}}]^T$,
the fermion number $\hat n_{i} \equiv \hat n_{i\uparrow} + \hat n_{i\downarrow}$,
the occupation projectors 
$\hat h_{i} \equiv|0\rangle\langle0|_{i}$ (hole) and
$\hat d_{i} \equiv
\left|\uparrow\downarrow\rangle\langle\uparrow\downarrow \right|_{i}
\equiv \hat n_{i\uparrow} \hat n_{i\downarrow}
$ (doublon), etc.
The spin-spin $\langle \hat S_{i} \cdot \hat S_{j}\rangle$ and 
hole-doublon $\langle \hat h_{i} \cdot \hat d_{j} \rangle$ 
correlations are schematically depicted in \Fig{Fig:XTRG}(a).

In our XTRG simulations, 
we consider the $L\times L$
square-lattice Hubbard model with $L=4,6,8$ with 
open boundary conditions, facilitating direct comparisons to 
optical-lattice measurements.
We also fully implement
non-Abelian spin and particle-hole (i.e., charge) symmetries. 
This allows us to reduce the $D$ states retained in XTRG
to an effective dimension of $D^\ast$ multiplets.
To be specific, for the half-filled 
case we exploit SU(2)$_{\rm charge} \,
\otimes$ SU(2)$_{\rm spin}$,
and for the doped case U(1)$_{\rm{charge}} \,
\otimes$SU(2)$_{\rm{spin}}$ symmetry.
In practice, this yields an effective
dimensional reduction of $D/D^\ast \sim 5.6$
and 2.6, respectively. This corresponds to
a $(D/D^\ast)^4 \simeq 30$-$1000$ fold 
reduction of computation time in the finite-$T$ simulations, guaranteeing
high efficiency and accuracy for fermion simulations. 
We obtain very well converged XTRG results on the $L=8$ 
square lattice at half filling (total site number $N=L^2=64$)
using up to $D^\ast=900$ multiplets ($D\simeq 5,000$ states),
and on the $L=6$ lattice upon doping using up to $D^\ast=1,200$ multiplets 
($D\simeq 3,100$ states)
\cite{SM} down to temperatures $T/t\simeq 0.06$ which 
is unprecedentedly low for such system sizes. 

The DQMC simulation performed here is of the finite 
temperature version with fast update~\cite{Assaad2008},
which has been successfully exploited in 
the finite-temperature simulation of 2D Hubbard model 
at half filling by some of the authors \cite{ChuangChen2019}.

\begin{figure}[tb]
\includegraphics[width=1\linewidth]{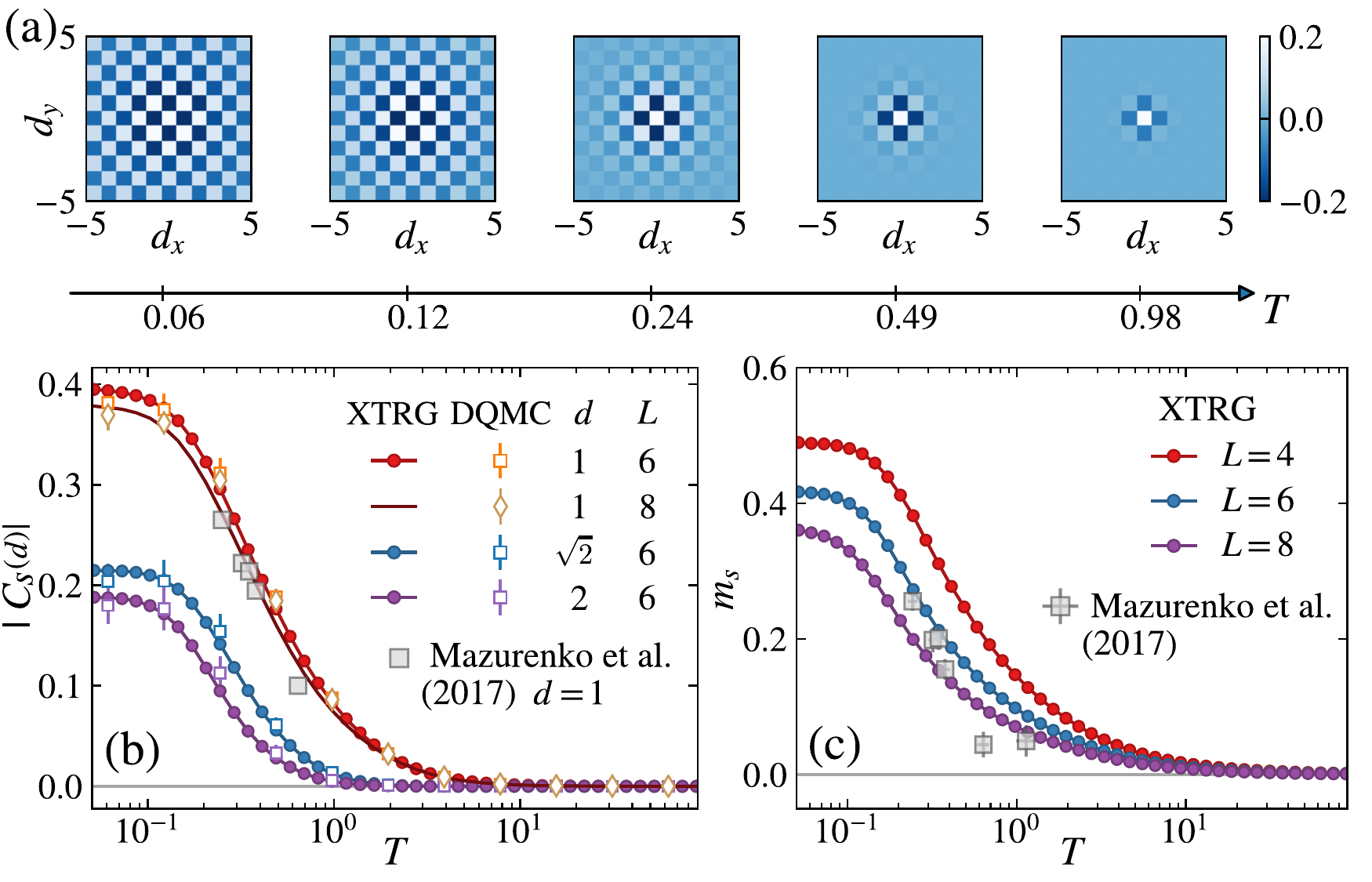}
\caption{Half-filled Fermi-Hubbard model with $U=7.2$ and sizes $L=4,6,8$.
(a) The finite-size AF order pattern is determined from the spin correlation $C_S(d)$
versus $(d_x, d_y)$, which melts gradually as $T$ increases. 
We show in (b) the spin correlation function $|C_S(d)|$ of various $d=1, \sqrt{2}, 2$,
and in (c) the finite-size spontaneous magnetization $m_s$ (see definition in the main text).
Excellent agreement between the calculated ($L=8$) data to the
experimental data \cite{Mazurenko2017} can be observed.
}
\label{Fig:Half}
\end{figure}

\begin{figure}[tbh]
\includegraphics[width=1\linewidth]{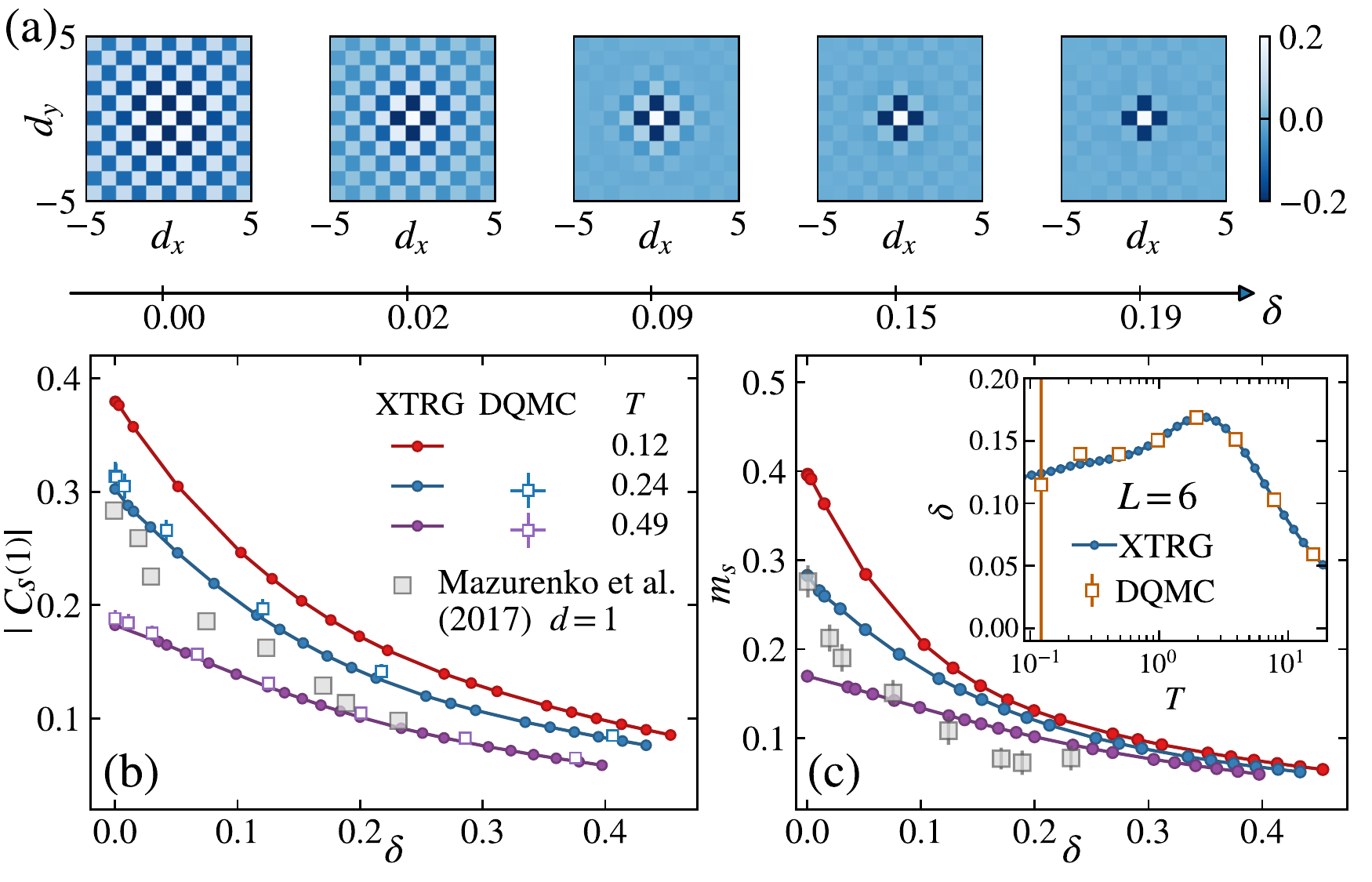}
\caption{Doped Fermi-Hubbard model with $U=7.2$ and size $L=6$. 
(a) shows the spin correlation pattern $C_S(d)$ versus doping $\delta$, 
plotted at the lowest temperature $T=0.06$, where the finite-size AF 
order fades out for $\delta \gtrsim 0.15$.
The computed (b) spin correlations $|C_S(d=1)|$
and (c) staggered magnetization $m_s$
are compared to the experimental data \cite{Mazurenko2017}.
The XTRG data in (b,c) are obtained via extrapolation $1/D^*\to0$ \cite{SM}.
In the inset of (c), we show how the doping $\delta$, 
computed by both XTRG and DQMC,
varies with $T$ at a fixed chemical potential $\mu=1.5$.
}
\label{Fig:Doped}
\end{figure}

\textit{Spin correlations and finite-size magnetic order at half-filling.---}
In recent experiments, the Fermi-Hubbard antiferromagnet (AF) 
has been realized in ultracold optical lattices at 
low effective temperature $T/t<0.4$ \cite{Mazurenko2017}. 
We first benchmark the XTRG method, along with DQMC, 
with the experimental results of the Fermi-Hubbard model at half-filing.
Fig.~\ref{Fig:Half}(a) reveals the finite-size AF magnetic structure 
by showing the spin-spin correlations 
$C_S(d) \equiv \frac{1}{N_d} \sum_{|{i} - {j}|=d}
\frac{\langle \hat S_{i} \cdot\hat S_{{j}} \rangle}{S(S+1)}$,
summed over all $N_d$ pairs of sites $i$ and $j$ with distance $d$,
where $i, j$ both denote 2D Cartesian coordinates for the sites in the original square lattice. 
The real-space spin structure shows AF magnetic order across the finite-size system at low 
temperature, e.g., $T\lesssim0.12$, which melts gradually as temperature 
increases. The AF pattern effectively disappears above 
$T \sim 0.49$, in good agreement with recent experiments \cite{Mazurenko2017}. 
In Fig.~\ref{Fig:Half}(b), we show $|C_S(d)|$ vs. $T$ at three fixed values of 
$d = 1, \sqrt{2}, 2$. Our XTRG and DQMC 
curves agree rather well in the whole temperature range,
for both $L=6$ and 8. \Fig{Fig:Half}(c) shows the spontaneous magnetization 
$m_s \equiv \sqrt{S(\pi,\pi)}$ vs. $T$ for $L=4,6,8$.
Here $S(q)= \frac{1}{N(N-1)} \sum'_{{i},{j}}
\frac{\langle \hat S_{i} \cdot\hat S_{{j}} \rangle}{S(S+1)}
e^{-\mathrm{i}{q}\cdot({i}-{j})}$
is the spin structure factor,
where the summation excludes on-site correlations
(following the convention from experiments \cite{Mazurenko2017}) and $N=L^2$ the total system size. 
For all system sizes considered, the spontaneous magnetization $m_s$ grows quickly 
as $T$ is decreased from $\simeq1$ to $\simeq0.1$. 
Notably, for both spin correlations and spontaneous magnetization, 
the $L=8$ XTRG data shows good qualitative agreement with the experimental  
measurements. This may be ascribed to the similar system sizes and boundary conditions, 
i.e., $8\times 8$ open square lattices vs. approximately 75-site optical lattice 
in experiments \cite{Mazurenko2017}.

\textit{Staggered magnetization upon hole doping.---}
By tuning the chemical potential $\mu < U/2$, 
we dope holes into the system and study how they affect 
the magnetic properties. \Fig{Fig:Doped}(a) shows 
the spin correlation patterns for different dopings $\delta$ at low $T$. 
The AF order clearly seen at low doping, becomes increasingly short ranged 
as $\delta$ increases, effectively reduced to nearest-neighbor (NN) only 
for $\delta \gtrsim 0.15$. The fall-off of AF order upon doping 
can also be observed in $|C_S(d)|$ with a fixed distance $d$. 
In Fig.~\ref{Fig:Doped}(b), we show the $d=1$ NN 
spin correlations. Our XTRG and DQMC
agree well for $T=0.49$ and 0.24, while the sign problem hinders
DQMC from reaching the lowest $T=0.12$ \cite{SM}.

\Fig{Fig:Doped}(c) shows the staggered magnetization $m_s$ vs. $\delta$.
Again a rapid drop of the finite-size AF order at approximately 
$\delta \in [0.1, 0.25]$ can be seen. 
The qualitative agreement with experimental measurements seen
in Fig.~\ref{Fig:Doped}(b,c)
suggest that the effective temperature 
of ultracold fermions falls between $T/t= 0.24$ and 0.49, 
consistent with the experiments \cite{Mazurenko2017}.
In our calculations we tune the doping 
$\delta$ by scanning the chemical potentials $\mu$. 
In the inset of \Fig{Fig:Doped}(c), we show the doping
$\delta$ vs. $T$ for a fixed $\mu=1.5$
(again the XTRG and DQMC results agree  
for $T\gtrsim 0.24$ with a tolerable sign problem \cite{SM} for DQMC).
The behavior of $\delta$ is non-monotonic:
it first increases as $T$ is lowered [having $\delta(T=\infty)=0$],
and then slowly decreases due to hole repulsion 
(see hole-hole correlation vs. $T$ in \cite{SM}).

\begin{figure}[tb]
\includegraphics[width=0.96\linewidth]{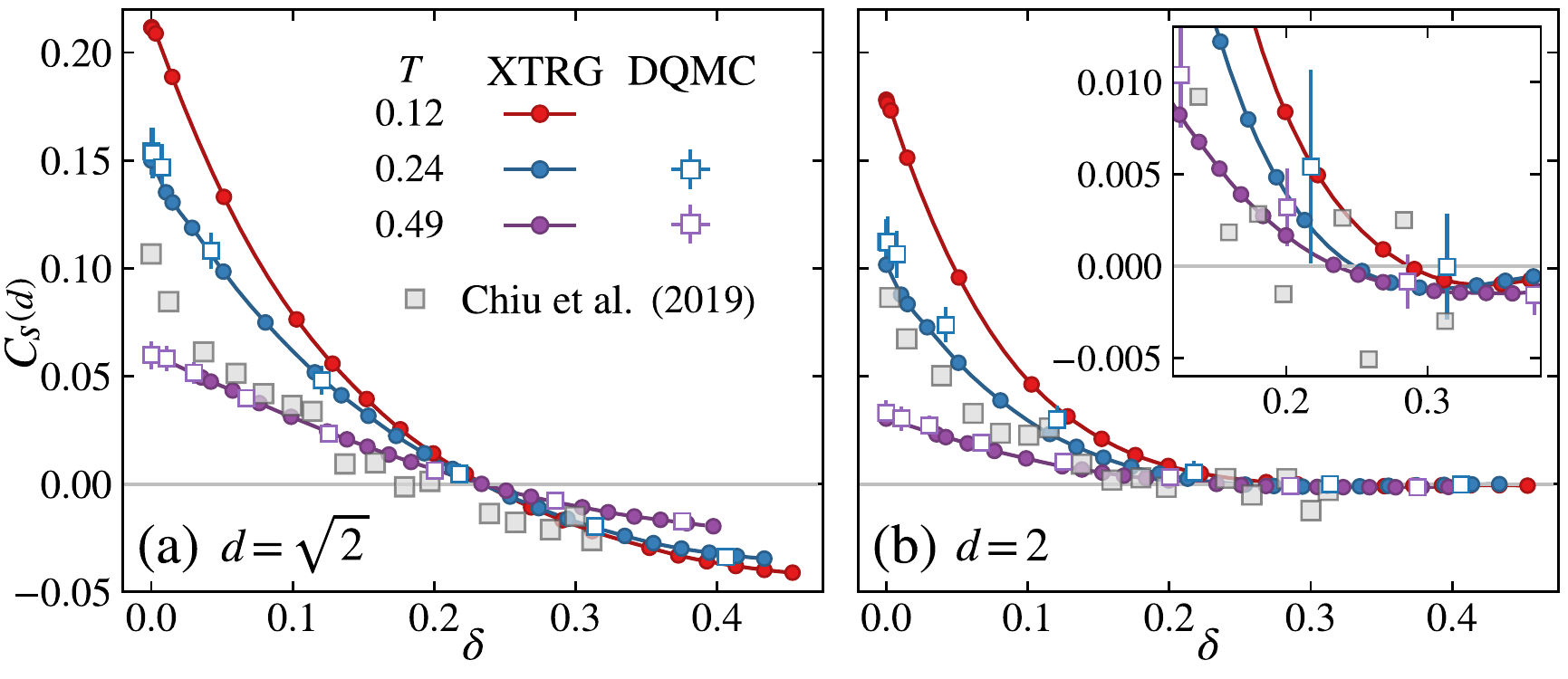}
\caption{
Diagonal and NNN $C_S(d)$ correlations as functions of doping $\delta$ for a 
$6\times6$ system with $U=7.2$ for (a) $d=\sqrt{2}$ and (b) $d=2$. 
The inset to (b) zooms in 
on small $C_S(d)$ values. The sign-reversal
of $C_d$ is in 
good agreement with experimental data \cite{Chiu2019}.
}
\label{Fig:SignChng}
\end{figure}

\textit{Magnetic polarons.---}
In \Fig{Fig:SignChng}, we analyze spin correlations 
between the diagonal ($d{=}\sqrt{2}$) and next-nearest-neighbor
($d{=}2$, NNN) sites. We compare them to 
recent measurements where it was 
found that the diagonal correlation 
$C_S(\sqrt{2})$ undergoes a sign reversal
around $\delta \simeq 0.2$ \cite{Chiu2019}.
Our computations reproduce this fact [\Fig{Fig:SignChng}(a)].
For the NNN correlations ($d=2$) [\Fig{Fig:SignChng}(b)], 
we find that an analogous sign reversal, hardly discernible 
in experiments, takes place around $\delta\simeq0.25$.

The sign reversal can be explained 
within the geometric string theory \cite{Grusdt2018}. It 
signals the formation of a magnetic polaron in the system.
As shown in Fig.~\ref{Fig:XTRG}(c),
the hole motion through the system generates a 
string of misaligned spins. The strong NN AF spin 
correlations are thus mixed with the diagonal and even further 
correlations, e.g. $C_S(2)$, resulting in even ferromagnetic clusters
[red and blue shaded regions in Fig.~\ref{Fig:XTRG}(c)]. 
Due to the interplay between the charge impurity and 
magnetic background, the moving hole distorts 
the AF background around the dopant 
[see the gray ``cloud" in Fig.~\ref{Fig:XTRG}(b)], 
giving rise to a collective excitation, i.e., the magnetic polaron.
Such exotic quasi-particles in doped Hubbard system
have been imaged experimentally \cite{Koepsell2019}
for a doublon in the particle-doped case. 

\begin{figure}[htb]
\includegraphics[width=1\linewidth]{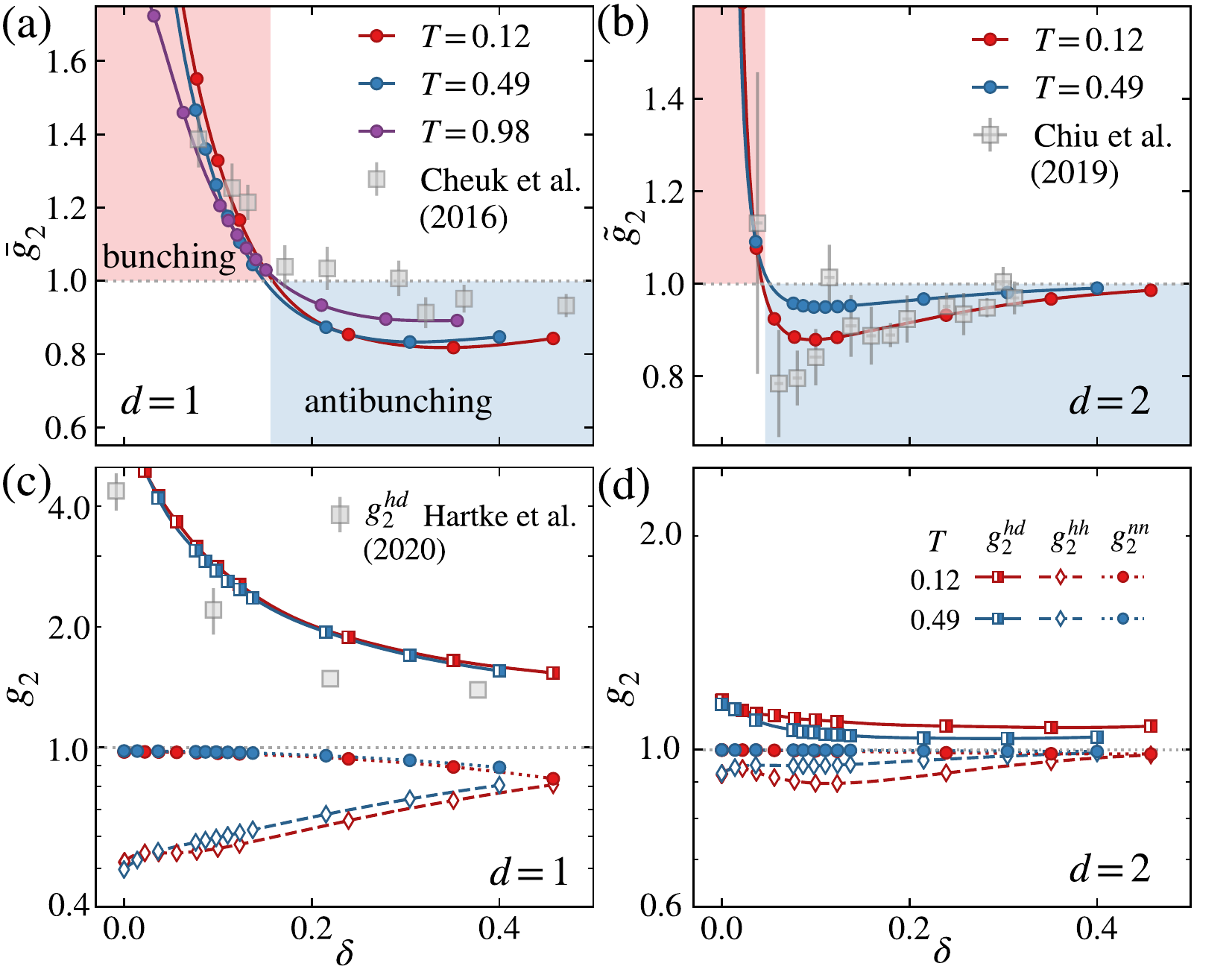}
\caption{ 
Various $g_2$ correlators for a $6\times6$ system with $U=7.2$. 
The antimoment correlators (a) $\bar g_2(d=1)$ and (b) $\tilde g_2(d=2)$
are shown as functions of doping $\delta$.
Experimental data with $d=1$, $T/t\simeq1.0$ \cite{Cheuk2016b} 
and $d=2$, $T/t\simeq0.25$ \cite{Chiu2019} are included for comparison.
(c, d) The two-cite hole-doublon ($g_2^{hd}$), hole-hole ($g_2^{hh}$), and full-density ($g_2^{nn}$)
correlations, for (c) $d=1$ and (d) $d=2$. The $d=1$ hole-doublon correlations $g_2^{hd}$
is compared with experiment in (c), 
where a nice agreement is observed despite a separate 
$U/t\simeq 11.8$ in experiment \cite{Hartke2020}.
}
\label{Fig:g2}
\end{figure}

\textit{Hole-doublon bunching and hole-hole antibunching.---} 
Quantum gas microscope can also access parity-projected {\it antimoment} correlation functions 
defined in the charge sector, 
$\bar g_2(d) \equiv \frac{1}{N_d} \sum_{|{i}-{j}|=d} \frac{\langle \hat \alpha_{i} \,
\hat \alpha_{j} \rangle}{\langle \hat \alpha_{i} \rangle\langle\hat \alpha_{j} \rangle}$ \cite{Cheuk2016b}
and 
$\tilde g_2 \equiv
\frac{1}{N_d}\sum_{|{i}-{j}|=d}\frac{1}{\delta^2}
[\langle \hat \alpha_{i}\hat \alpha_{j}\rangle 
- \langle\hat \alpha_{i}\rangle\langle\hat \alpha_{j}\rangle]$ \cite{Chiu2019},
with the antimoment projector 
 $\hat{\alpha}_{i}
\equiv \hat h_{i} + \hat d_{i}$
\footnote{
A local (spin) moment is present only at filling
$n=1$. An analogous moment can be defined in
the SU(2) particle/hole sector, which is complimentary to
the spin space as it operates within empty and double
occupied state, hence `antimoment'
}.
\Fig{Fig:g2}(a,b) shows the computed antimoment correlation results.
Antimoments are bunching ($\bar{g}_2>1$) at low doping, 
yet become antibunching ($\bar{g}_2<1$)  at large doping, 
in quantitative agreement with an earlier $^{40}$K 
experiment \cite{Cheuk2016b} and a more recent $^6$Li gas 
measurement \cite{Chiu2019}, see Figs.~\ref{Fig:g2}(a) and (b), respectively.
The antibunching at large doping is attributed to hole repulsion, 
and the bunching at low-doping to hole-doublon pairs \cite{Cheuk2016b}. 

Now antimoments contain contributions
from both, holes and doublons, yet their individual 
contributions cannot be distinguished via  
parity projection measurements \cite{Cheuk2016b,Chiu2019}. 
XTRG, however, readily yields detailed correlators
$g^{ll'}_2(d) 
\equiv \frac{1}{N_d} \sum_{|\mathbf{i-j}|=d} \tfrac{\langle \hat l_{i} \, \hat l'_{{j}}\rangle}
{\langle \hat l_{i} \rangle\langle\hat l'_{j}\rangle}$,
with $l \in \{h,d\}$ and
$\hat{l}_i\in\{ \hat h_{i}, \hat{d}_{i} \}$
for hole or double-occupancy projectors, respectively.
Later we also use $l=n$ for $\hat{l}_j = \hat{n}_j$
the local density.

Our results for the correlations $g_2^{hh}(d)$ and $g_2^{hd}(d)$
vs. $\delta$ are shown in \Fig{Fig:g2}(c,d). We always find 
$g_2^{hh}(d)<1$ and therefore anticorrelation amongst holes,
while $g_2^{hd}>1$ corresponds to strong bunching between holes and doublons.
As shown in \Fig{Fig:g2}(c), the computed $g_2^{hd}$ data show remarkable
agreement with very recent experimental measurements using the full-density-resolved
bilayer readout technique \cite{Hartke2020,Koepsell2020}. The change from bunching to antibunching
behaviors in antimoment correlations in Fig.~\ref{Fig:g2}(a,b) can be ascribed to the fact that
the hole-doublon attraction is advantageous over the hole-hole repulsion at low doping 
while the latter dominates at relatively large doping \cite{SM}. 
When comparing the charge correlations at $d=1$ and 2 in Fig.~\ref{Fig:g2}(c,d), 
we find that the hole-doublon bunching effect in $\bar g_2(1)$ is particularly strong at $\delta \ll 1$, 
where the holes mostly stem from NN hole-doublon pairs [see illustration in Fig.~\ref{Fig:XTRG}(b)].
The further-ranged $g_2^{hd}(2)$ still shows the bunching effect, yet gets much reduced.

The full density correlation $g_2^{nn}(d)$ is shown in \Fig{Fig:g2}(c, d). 
We observe $g_2^{nn}(d) \approx 1$ 
at low doping for both $d=1,2$, i.e., weak non-local charge correlations 
near half-filling, and a more pronounced 
anti-correlation $g_2^{nn}(d) < 1$ as $\delta$ increases.
Based on our XTRG results, we further reveal that the
longer-ranged $g_2^{nn}(2)$ also exhibits anticorrelations upon doping, 
suggesting the statistical Pauli holes may be rather nonlocal,
though decaying rapidly spatially.

\textit{Conclusion and outlook.---} 
In this work, we generalized XTRG \cite{Chen2018,Li2019} to the 2D Fermi-Hubbard model. 
Employing XTRG and DQMC, we obtained reliable results both for half-filling 
and doped cases and found consistency with the
ultracold atom experiments. 
XTRG can explore a broader parameter space, especially in the doped case, than DQMC, which is 
limited by a minus-sign problem. 
XTRG$+$DQMC constitutes a state-of-the-art complimentary numerical setup 
for probing the phase diagram of Fermi-Hubbard models, 
for SU(2) fermions here and generally SU(N) fermions \cite{Takahashi2018}, 
thanks to the implementation of non-Abelian symmetries \cite{Weichselbaum2012}.
Fundamental questions, such as the explanation of the Fermi arcs and the
pseudogap phase~\cite{Norman1998,Keimer2015}, with their implications 
for the breaking of Luttinger's theorem~\cite{Luttinger1960,Oshikawa2000,Paramekanti2004,Senthil2003}, 
or the role of topological order~\cite{Gazit2019,ChuangChen2020,ChuangChen2020b}
are open interesting topics to 
be studied by XTRG$+$DQMC and optical lattices.

\textit{Acknowledgments.---}
B.-B.C. and C.C. contributed equally to this work.
The authors are greatly indebted to Fabian Grusdt and Annabelle Bohrdt for 
numerous insightful discussions.
W.L., J.C., C.C. and Z.Y.M. are supported by 
National Natural Science Foundation of China
(Nos. 11974036, 11834014, 11921004, 11904018) and the Fundamental Research Funds 
for the Central Universities. Z.Y.M. is also supported by the RGC of Hong Kong SAR China (Grant Nos. 17303019 and 17301420). 
The German Research Foundation (DFG) supported this research through WE4819/3-1 (B.-B.C.) and Germany's Excellent Strategy -- EXC-2111 -- 390814868. 
A.W. was supported by the U.S. Department of Energy, 
Office of Basic Energy Sciences, under Contract No. DE-SC0012704.
We thank the Center for Quantum Simulation Sciences in the Institute of Physics, 
Chinese Academy of Sciences, the Computational Initiative at the 
Faculty of Science at the University of Hong Kong, the Tianhe platforms at the National 
Supercomputer Centers in Tianjin and Guangzhou, 
and the Leibniz-Rechenzentrum in Munich for their
technical support and generous allocation of CPU time.

\AtEndEnvironment{thebibliography}{
\bibitem{SM} In Supplementary Materials, we briefly recapitulate the basic idea of XTRG and
its generalization to interacting fermions. The implementation of non-Abelian symmetries is 
provided in {\color{blue}Sec.~A}. Detailed convergence check and the linear extrapolation $1/D^*\to0$ 
for the spin correlation are shown in {\color{blue}Sec.~B}. 
Complementary XTRG data on the spin and charge correlations in the doped Hubbard model 
are presented in {\color{blue}Sec.~C}, and {\color{blue}Sec.~D} 
is devoted to details of DQMC algorithms and calculations.
}
\bibliography{fXTRG.bib}

\begin{center}
\textbf{\large Supplementary Materials: Quantum Many-Body Simulations of the 2D Fermi-Hubbard Model in Ultracold Optical Lattices}
\end{center}
\setcounter{equation}{0}
\setcounter{figure}{0}
\setcounter{table}{0}
\setcounter{section}{0}
\setcounter{page}{1}
\setcounter{section}{0}
\makeatletter
\renewcommand{\theequation}{S\arabic{equation}}
\renewcommand{\thefigure}{S\arabic{figure}}
\renewcommand{\theequation}{S\arabic{equation}}
\renewcommand{\thesection}{\Alph{section}}
\renewcommand{\thesubsection}{\arabic{subsection}}

\section{A. Exponential Tensor Renormalization Group Approach for Correlated Fermions}
\label{SM:fXTRG}

\begin{figure}[thb]
\includegraphics[width=1\linewidth]{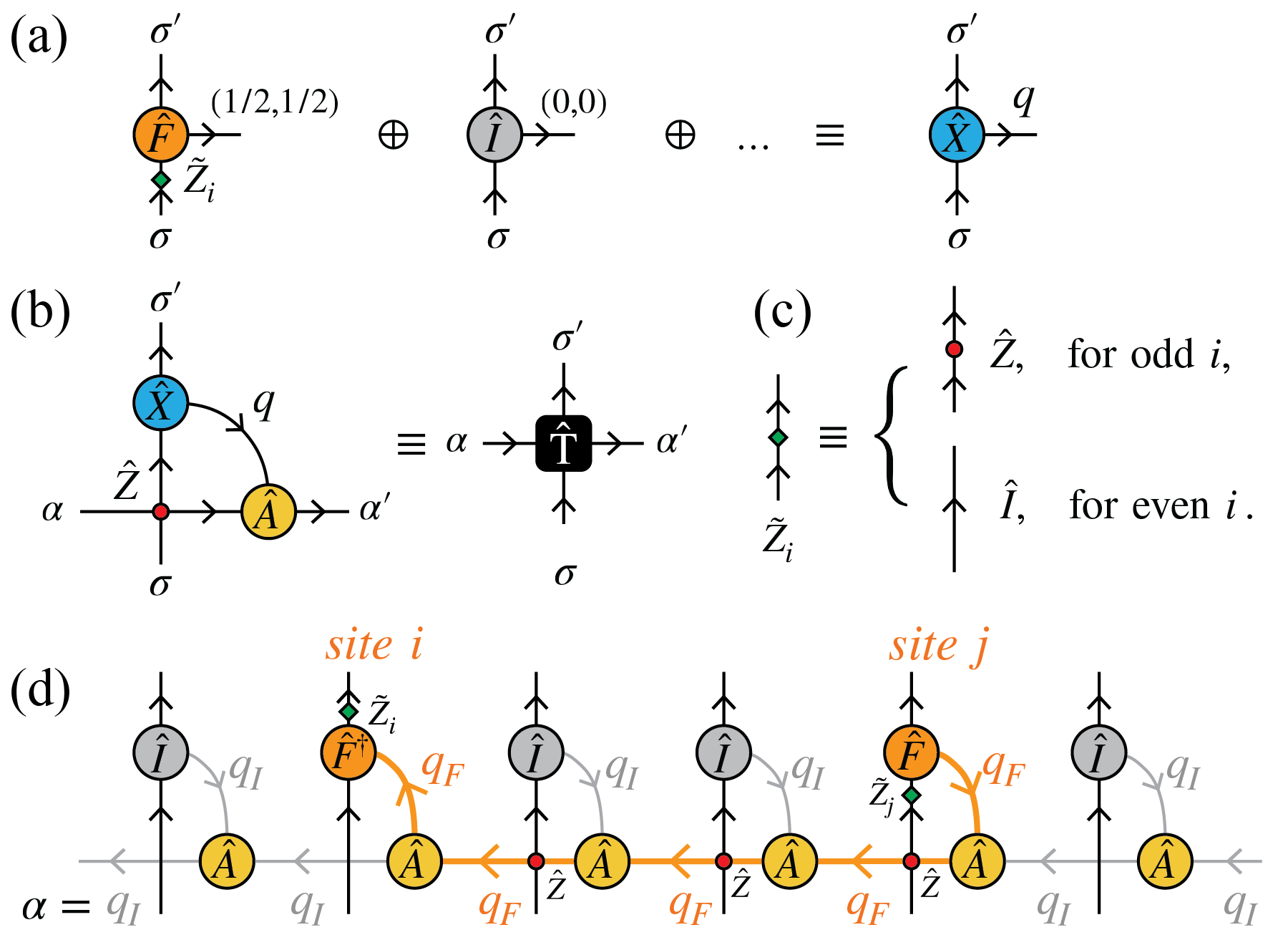}
\caption{\textit{Tensor-network representation for fermion operators. ---}
(a) An irreducible operator (irop) can always be
assigned an irop index, shown as the
horizontal line sticking out towards the right
of a tensor indicated by a circle. The vertical
lines describe a local state space.
The irop index is also assigned 
symmetry labels $q\equiv (C,S)$
which describe the transformation of the operator
under given charge ($C$) and spin ($S$) symmetry.
Here examples of local irops are 
the fermion operator $\hat{F}$ 
[with $q_F \equiv (1/2,1/2)$] or a trivial identity
operator $\hat I$ [with $q_I\equiv(0,0)$].
These may be combined into the (non-irop)
tensor $\hat{X}$ that now describes a (to the extent
required) complete local operator basis.
(b) For the MPO of the Hamiltonian,
the local tensor $T$ is
constructed from the local operator basis $\hat X$
and the $A$ tensor of a super-MPS, connected 
by the operator basis indexed by $q$. 
Fermionic signs are taken are of by
the charge parity operator $\hat{Z}\equiv (-1)^{\hat{n}}$ 
which needs to be applied at every 
crossing point of lines if negative charge parity
can occur on both lines
(this is completely analogous, e.g., to the swap gate
in fermionic iPEPS \cite{Corboz2010}).
It is denoted by the small red dot.
(c) When using SU(2) particle-hole symmetry,
the local fermion operator is decorated with 
an additional $\tilde Z_i$ 
(denoted by the green diamond), $\hat{F}_i \to \tilde Z_i \hat{F}_i$,
to recover the correct hopping structure in terms of signs,
with $\tilde Z_i \in \{ \hat{I}, \hat{Z} \}$ 
for even (odd) sites $i$, respectively.
(d) A single hopping term in the Hamiltonian, i.e.,
$h_{i,j}\equiv \hat{F}_i^\dagger \cdot \hat{F}_j^{\ }$
from site $j$ to site $i$
is constructed in MPO form from the local tensor
as schematically depicted 
in panel (b). Local terms in the Hamiltonian are also
added to the local MPO basis $\hat{X}$ [suggested by the
$\oplus \ldots$ in (a)], e.g., with the onsite interaction
given by  $(\hat{n}_{i} - 1)^2\equiv \tfrac{4}{3} \hat{C}_i^\dagger
\cdot \hat{C}_i^{\ }$, i.e., the Casimir operator
in the SU(2)$_\mathrm{charge}$ symmetry.
}
\label{Fig:fMPO}
\end{figure}

\subsection{1. Renormalization group algorithms for 2D fermion models}
Renormalization group numerical methods provide powerful tools tackling 
fermion many-body problems. Among others,
the density-matrix \cite{Noack1994} and tensor-network 
renormalization group (TRG) \cite{Corboz2010,Kraus2010,Gu2010} 
methods have been developed to simulate fermion models in two dimensions (2D), 
with focus on the $T=0$ properties, playing an active role in solving the challenging
Fermi-Hubbard model at finite doping \cite{LeBlanc2015,Zheng2017,Qin2019,Chung2020}.

For $T>0$, thermal TRG algorithms exploits the purification framework 
in simulating thermodynamics of both infinite- and finite-size systems 
\cite{Verstraete2004,Zwolak2004,Feiguin2005,Li2011}. 
Recently, generalizations of DMRG-type calculations to finite
temperature have become available via matrix-product-state samplings \cite{Bruognolo2017,Chung2019}
and the exponential TRG (XTRG) \cite{Chen2018,Chen2019,Li2019}. 
Most of the thermal TRG methods mainly apply to the spin/boson systems,
and there are few attempts for fermions at finite temperature.
For example, an infinite TRG approach has been proposed to simulate 
2D fermion lattice models directly in the thermodynamic limit, 
however it is restricted to relatively high temperature \cite{Czarnik2014}.
Therefore, it is highly desirable to have reliable and accurate 
TRG algorithms for simulating large-scale correlated fermion systems 
down to low temperatures.

XTRG can be employed to simulate large-scale system sizes,
e.g., width-8 cylinders for the square-lattice Heisenberg model \cite{Li2019}, and
width-6 cylinders \cite{Chen2019} for the triangular-lattice Heisenberg model, 
providing full and accurate access to various thermodynamic quantities 
as well as entanglement and correlations down to low temperature. 
Here, we generalize XTRG to 2D fermion models and perform the 
calculations on $L\times L$ open square lattices 
up to size $L=8$ (half filling) and $L=6$ (finite doping).

\subsection{2. Particle-hole and spin symmetries}

In the XTRG calculations of the Fermi-Hubbard model,  
we implement non-Abelian/Abelian particle-hole and spin symmetries 
in the matrix-product operator (MPO) representation of the Hamiltonian 
and the thermal density operators, which greatly reduces the computational
resources and makes the high-precision low-temperature simulations 
possible in XTRG. 
Here the symmetry implementation
is based on the QSpace tensor library \cite{Weichselbaum2012}.

To be specific, consider the SU(2)$_{\rm charge} \otimes$ SU(2)$_{\rm spin}$
symmetry as an example.
The SU(2)$_{\rm charge}$, i.e., particle-hole
symmetry is present in the Fermi-Hubbard
model at half-filling on a bipartite lattices, such as 
the square lattice considered in this work.
QSpace permits to turn symmetries on or off at will,
such that either of the symmetries above can also be reduced
to smaller ones, such as U(1)$_{\rm charge}$  or U(1)$_{\rm spin}$.
This is required for example in the presence of a chemical
potential or an external magnetic field, respectively.
Throughout, we stick here to the order convention that the
charge label comes first, followed by the spin label,
i.e., $q=(C,S)$. For SU(2)$_{\rm charge}$, the `$S_z$'
label corresponds to $\tfrac{1}{2}(n_i-1)$, that is,
one half the local charge relative to half-filling.

The fermion operators can be organized into an irreducible 
four-component spinor 
\cite{Weichselbaum2012}, 
\begin{equation}       
\hat{F}_{{i}}^{(1/2,1/2)} = 
  \left(
  \begin{array}{c}
          s_{i} \hat{c}^{\dagger}_{i\uparrow} \\
             \hat{c}_{i\downarrow} \\
          s_{i} \hat{c}^{\dagger}_{i\downarrow} \\
          - \hat{c}_{i\uparrow}        
 \end{array}
 \right)\text{ .}
\label{Eq:FCS}        
\end{equation}
It is an irop that transforms like $q_F=(1/2,1/2)$.
Because it consists of multiple components, this
results in the third index [depicted as leg to the right in
Fig.~\ref{Fig:fMPO}(a)].
The local Hilbert space $\sigma_{(i)}$ of a site $i$
with $d=4$ states can be reduced to $d^\ast=2$ multiplets,
$q_\sigma=(1/2,0)$
combining empty and double occupied, i.e., hole and double
states, and $q_\sigma=(0,1/2)$ for the local spin
$S=1/2$ multiplet at single occupancy.

In Eq.~(\ref{Eq:FCS}), the index $i\equiv (i_1,i_2)$ denotes a 
2D Cartesian coordinate of the site in original square lattice. 
The implementation of SU(2)$_{\rm charge}$ requires a
bipartite lattice, $\mathcal{L} = \mathcal{A} \cup \mathcal{B}$,
which we distinguish
by the parity $s_i=\pm 1$, e.g., choosing arbitrarily but
fixed that the sites in $\mathcal{A}$ are even,
i.e., have $s_i=+1$ for $i \in \mathcal{A}$.
In practice, we adopt a snake-like mapping of the 2D square lattice 
(as shown in Fig.~\ref{Fig:XTRG}), with a 1D site ordering index $i$.
This leads to a simple rule: a site with $i \in$ even (odd) 
site of the quasi-1D chain also corresponds to the even (odd) 
sublattice of the square lattice with $s_i = \pm 1$.

For  SU(2)$_{\rm charge}$, 
to recover the correct hopping term in the Hamiltonian, 
this requires the alternating sign factor $s_{i}$. 
In fact, this alternating sign 
can be interpreted as 
different fermion orderings
on the even and odd sites \cite{Weichselbaum2012}, i.e., 
\begin{eqnarray}
  | \updownarrows \rangle_i & = & s_i \ \hat c^\dagger_{i\uparrow} \hat c^\dagger_{i\downarrow} | 0 \rangle =
  \begin{cases}
         \, \hat c^\dagger_{i\downarrow} \hat c^\dagger_{i\uparrow} | 0 \rangle, &\text{$i \in$ odd, $s_i=-1$,}\\
         \\
         \, \hat c^\dagger_{i\uparrow} \hat c^\dagger_{i\downarrow} | 0 \rangle, &\text{$i \in$ even, $s_i=1$.}
  \end{cases}
\label{eq:forder}
\end{eqnarray}
By reversing the fermionic order of every other site for the local state space as above,
we thus recover the correct structure in the electron hopping term 
\begin{equation}
\hat h_{i,j}  = \hat{F}_{i} 
^\dagger \cdot \hat{F}_{j}^{\ } 
= (\hat c_{i\uparrow}^\dagger \hat c_{j\uparrow}^{\ } + \hat c_{i\downarrow}^\dagger \hat c_{j\downarrow}^{\ }) + \mathrm{H.c.},
\label{Eq:hoppingterm}
\end{equation}
with site $i$ and $j$ always belonging to different sublattices of the square lattice.
By summing over all pairs of hopping terms, we recover
the tight-binding (TB) kinetic energy term 
on the square lattice, whose Hamiltonian reads
\begin{equation}
\hat H_{\text{TB}} = \sum_{\langle i,j \rangle} \hat h_{i,j}
= \sum_{\langle i,j \rangle}
  \hat{F}_{ i} 
  ^\dagger \cdot \hat{F}_{j}^{\phantom{\dagger} } 
\text{ .}\label{Eq:TBMPO}
\end{equation}
By the structure of a scalar product,
\Eq{Eq:TBMPO} explicitly reveals 
the SU(2) particle-hole and spin symmetry. 

When the interaction $U$ is turned on, the Fermi-Hubbard 
Hamiltonian [see \Eq{Eq:Hub} in the main text] remains
SU(2)$_{\rm{charge}} \otimes$SU(2)$_{\rm{spin}}$ invariant, 
as long as half-filling is maintained, i.e., $\mu=U/2$.
Then $$\sum_i U \hat n_{i\uparrow} 
\hat n_{i\downarrow} -\tfrac{U}{2}  (\hat n_{i\uparrow} + \hat n_{i\downarrow})
\equiv \tfrac{U}{2}  \sum_i (\hat{n}_i-1)^2 +\text{const.} $$
has a SU(2) charge symmetry, and the system has a totally symmetric 
energy spectrum centered around $C_z=0$. It is 
proportional to the Casimir operator of SU(2)$_\mathrm{charge}$.
However, when $\mu \neq U/2$, this acts like a magnetic
field in the charge sector, and
the SU(2)$_{\rm{charge}}$ symmetry is reduced to 
U(1)$_{\rm{charge}}$.

\subsection{3. Fermionic MPO}

Given this symmetric construction of the local fermionic operator $\hat{F}_i$ 
we describe below how to represent the many-body Hamiltonian as a fermionic MPO,
by taking the square-lattice tight-binding model Eq.~(\ref{Eq:TBMPO}) 
mentioned above as an example.
We first introduce a super matrix product state (super-MPS) representation in Fig.~\ref{Fig:fMPO}, 
which encode the ``interaction" information compactly and can be conveniently 
transformed into the MPO by contracting the $A$ tensor 
with the local operator basis $\hat X$, 
as shown in Fig.~\ref{Fig:fMPO}(a,b).

\begin{table}[]
\begin{tabular}{lllll}
\hline
\multicolumn{1}{|c|}{$\Vert A_{\alpha, \alpha'}^{q\, [k]}\Vert $} 
& \multicolumn{1}{l|}{$\alpha$} & \multicolumn{1}{l|}{$\alpha'$} & \multicolumn{1}{l|}{$q$} & \multicolumn{1}{l|}{$k$}  \\ \hline
\multicolumn{1}{|c|}{1.} & \multicolumn{1}{l|}{(0,0)}  & \multicolumn{1}{l|}{(0,0)}  & \multicolumn{1}{l|}{$(0,0)$} &  \multicolumn{1}{l|}{$k<i$ or $k>j$} \\ 
\multicolumn{1}{|c|}{1.} & \multicolumn{1}{l|}{(0,0)}  & \multicolumn{1}{l|}{$(\tfrac{1}{2},\tfrac{1}{2})$}  & \multicolumn{1}{l|}{$(\tfrac{1}{2}, \tfrac{1}{2})$} &  \multicolumn{1}{l|}{$k=i$} \\ 
\multicolumn{1}{|c|}{1.} & \multicolumn{1}{l|}{$(\tfrac{1}{2},\tfrac{1}{2})$}  & \multicolumn{1}{l|}{(0,0)}  & \multicolumn{1}{l|}{$(\tfrac{1}{2}, \tfrac{1}{2})$} &  \multicolumn{1}{l|}{$k=j$} \\ 
\multicolumn{1}{|c|}{1.} & \multicolumn{1}{l|}{$(\tfrac{1}{2},\tfrac{1}{2})$}  & \multicolumn{1}{l|}{$(\tfrac{1}{2},\tfrac{1}{2})$}  & \multicolumn{1}{l|}{$(0, 0)$} &  \multicolumn{1}{l|}{$i<k<j$} \\[1ex] \hline
\end{tabular}
\caption{The nonzero reduced tensor elements
$\Vert A_{\alpha, \alpha'}^{q\, [k]}\Vert$
at site $k$ [cf. Fig.~\ref{Fig:fMPO}(d)], 
in the MPO representation of a specific hopping term $h_{i,j}$ (Eq.~\ref{Eq:hoppingterm}) between site $i$ and $j$.
The indices $\alpha, \alpha'$, and $q$ are labeled by symmetry
quantum numbers $(C,S)$.}
\label{Tab:RedTenA}
\end{table}

To be specific, consider a single hopping
term $h_{i,j}$ between site $i\neq j$ in Fig.~\ref{Fig:fMPO}(d). 
In the super-MPS, the corresponding
$A$ tensors have a simple internal structure, as
listed in Tab.~\ref{Tab:RedTenA},
since the main purpose of the $A$-tensor is to {\it route}
lines through. Hence they also contain simple Clebsch Gordan
coefficients, with the fully scalar representation
$(0,0)$ always at least on one index.
In $\Vert A_{\alpha, \alpha'}^{q\, [k]}\Vert$, 
$\alpha,\alpha'$ can be
$q_\mathrm{F}=(\tfrac{1}{2},\tfrac{1}{2})$ 
or $q_\mathrm{I} = (0,0)$ 
as shown in 
Fig.~\ref{Fig:fMPO}(c). Correspondingly, 
this contracts with either the fermion operator
$\hat{F}$ or $\hat{I}$ in $\hat{X}^q$, respectively. 
Contracting  $\hat{X}^q$ onto $A$, 
this casts the super-MPS which is made of $A$-tensors only, into 
``MPO" form consisting of the rank-4 tensors $T$,
as indicated in Fig.~\ref{Fig:fMPO}(b).
With the index $\alpha$ {\it routed} from site $i$
to site $j$, its $q$-label is fixed to that of the irop.
Therefore each single hopping term $h_{i,j}$ 
can be represented as an 
MPO as in Fig.~\ref{Fig:fMPO}(d), 
with reduced bond dimension $D^*=1$ (one multiplet per geometric bond).
Following a very similar procedure as in XTRG for spin systems \cite{Chen2018}, 
we can thus sum over all $h_{i,j}$ terms and obtain a compact MPO 
representation of the Hamiltonian Eq.~(\ref{Eq:TBMPO})
through variational 
compression which as part of the initialization is cheap.
This guarantees that an MPO with minimal 
bond dimension $D^\ast$ is obtained.

\subsection{4. Fermion parity operator $\hat Z$}

The $\hat{X}^q$ operator basis acts on the local fermionic Hilbert space, 
and thus fermionic signs need to be accounted for 
in the construction 
of the MPO representation of the Hamiltonian. 
As shown in Fig.~\ref{Fig:fMPO}(d), we introduce a product of 
parity operators $\hat Z$ between site $i$ and $j$,
generating a Jordan-Wigner string connecting the
operators $\hat{F}_{i}^\dagger$
and $\hat{F}_{j}$.
The parity operator $\hat Z$ is defined as $(-1)^{2C+1}$
for any state space, which yields $z=+1$ if $C$ is half-integer
(e.g., $C=1/2$ for empty and double occupied), and $z=-1$,
otherwise (e.g. $C=0$ for a singly occupied site).
In practice, for SU(2)$_\mathrm{charge}$,
based on \Eq{Eq:FCS} we use for even sites ($s_i=+1$)
\begin{equation}       
\hat{F} \equiv \hat{F}_{\text{even}}^{(1/2,1/2)} = 
  \left(
  \begin{array}{c}
            \hat{c}^{\dagger}_{i\uparrow} \\
            \hat{c}_{i\downarrow} \\
            \hat{c}^{\dagger}_{i\downarrow} \\
          - \hat{c}_{i\uparrow}           
 \end{array}
 \right),
\label{Eq:FCSeven}        
\end{equation}
while for odd sites, we use (purely in terms of matrix
elements) in the MPO, $\hat{F}_\mathrm{odd} =
\hat{Z}\hat{F}$, instead
(cf. discussion with \Eq{eq:forder}; \cite{Weichselbaum2012}),
with the Hermitian conjugate $(\hat{Z}\hat{F})^\dagger =
\hat{F}^\dagger \hat{Z}$.
Therefore introducing $\tilde{Z}_i \in \{ \hat{I}, \hat{Z} \}$ 
for even (odd) sites $i$, respectively,
this takes care of the alternating sign structure,
as illustrated in Fig.~\ref{Fig:fMPO}(c),
and consistent with
\Eq{Eq:FCS}.

Overall, assuming $i<j$ with similar Fermionic order
in that site $i$ is added to the many body state space before site $j$,
the hopping term $\hat h_{i,j}$ can thus be represented as
\begin{equation}
\hat h_{i,j} =  (\hat{F}^\dagger \tilde{Z})_i
\otimes \hat Z_{i+1}  \otimes ... \otimes \hat{Z}_{j-1}
\otimes (\hat{Z} \tilde{Z} \hat{F})_j.
\end{equation}
Given the bipartite lattice structure, therefore
up to the dagger, the same $\hat{F}$ (or $\hat{Z}\hat{F}$)
is applied at both sites $i$ and $j$ depending on whether
$i$ is even (or odd), respectively.

\subsection{5. Exponential cooling and expectation values}

XTRG requires the MPO of the Hamiltonian as input for
initialization. Therefore when building the MPO for the
Hamiltonian, this is the only place where fermionic signs
play a role. Thereafter XTRG follows an automated machinery.
We
compute the thermal density operator $\hat \rho(\beta/2)$,
and then estimate thermodynamics quantities, entanglement, and correlations from it.
We start with a very high-$T$ density operator $\hat \rho_0(\tau)$ 
at inverse temperature $\tau \ll 1$, obtained via the series expansion \cite{Chen2017}
$$
\hat \rho_0(\tau) = \sum_k \tfrac{(-\tau)^k}{k!} \hat H^k.
$$
Here the initial $\tau$ can be exponential small, which
thus limits the series expansion to very few terms to already
reach machine precision for the initial $\hat \rho_0(\tau)$.
Then, we cool down the system exponentially
by squaring the density matrix. The $n$-th XTRG iteration 
yields
\begin{equation}
\hat \rho_{n-1}(2^{n} \tau) \otimes \hat \rho_{n-1}(2^{n} \tau) \rightarrow \hat \rho_n(2^{n+1} \tau).
\label{Eq:Cooling}
\end{equation}
With $\hat \rho_n(2^{n+1} \tau)$, we can compute thermal expectation values at
inverse temperatures $\beta_n = 2^{n+2} \tau$ using the thermofield double 
trick of purification \cite{Feiguin2005,Dong2017,Chen2018},
equivalent to the simple procedure in \Fig{Fig:XTRG}.

One advantage in the fermion XTRG is its simplicity.
In the cooling step $\hat \rho_{n-1} \otimes \hat \rho_{n-1} \rightarrow \hat \rho_{n}$ 
in Eq.~(\ref{Eq:Cooling}), no fermion parity operators $\hat{Z}$ are involved 
when we perform MPO iteration 
and compression just as for spin/boson systems. 
Besides, in the calculations of density-density correlations such as 
spin-spin and hole-hole(-doublon) correlations,
the charge quantum numbers $C$ in the $q$-label of operators $\hat S$ 
and $\hat{h}$ ($\hat{d}$) are always even, and thus the Jordan-Wigner 
string consists of trivial identity operators and can also be safely
ignored, as illustrated in Fig.~\ref{Fig:XTRG}(a) of the main text.
Even though not required here, also fermionic
correlations can be computed within fermionic XTRG, and proceeds
completely analogous to fermionic MPS expectation values, then
also with a Jordan Wigner string stretching in between sites
$i$ and $j$.

\begin{figure}[tb]
\includegraphics[width=\linewidth]{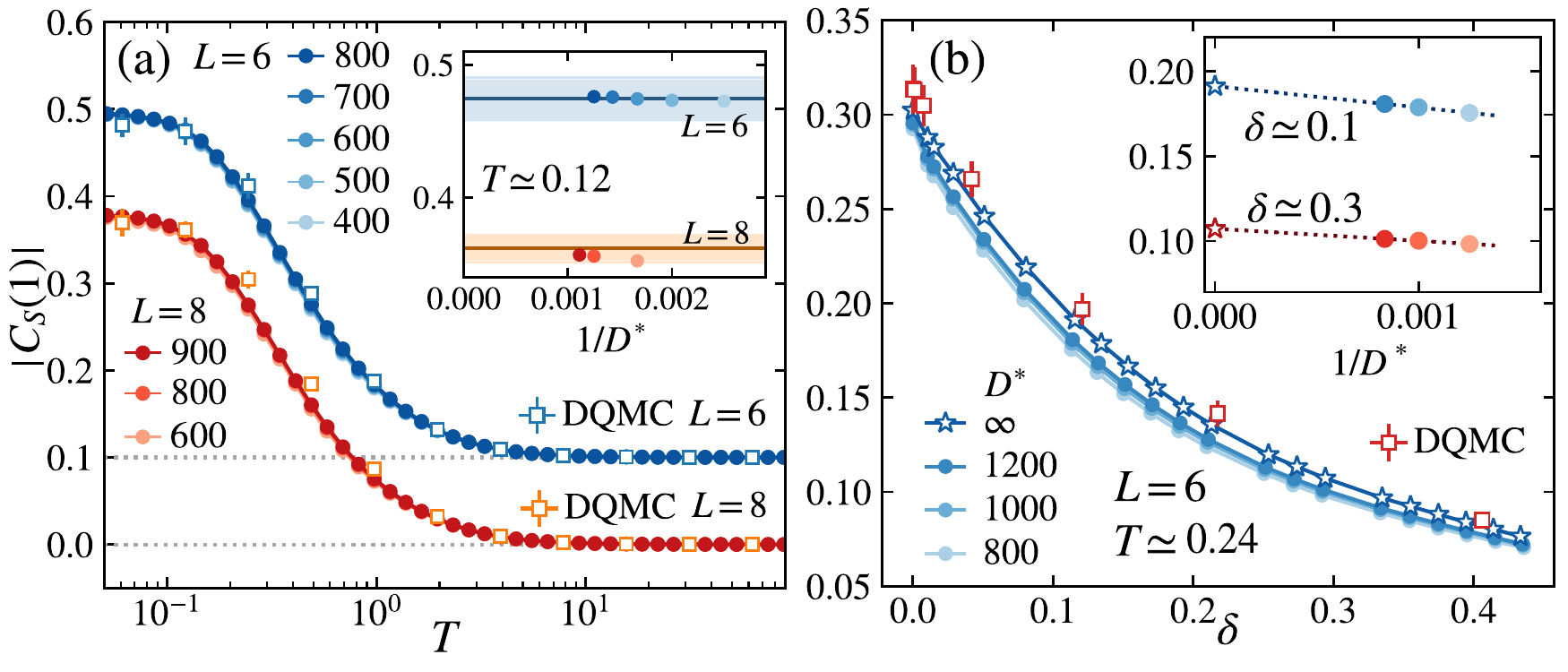}
\caption{\textit{XTRG$+$DQMC bechmark results.} 
(a) NN spin correlation $|C_S(d=1)|$ of half-filled square-lattice Hubbard system 
for $U=7.2$ and sizes $L=6, 8$, with $D^\ast=400$-$900$. 
The $L=6$ data have been shifted upwards by 0.1, for the sake of readability.
In the inset, $|C_S|$ at low temperature $T\simeq0.12$ is shown versus $1/D^\ast$, 
with the DQMC results [mean (line) and standard deviation (color matched shaded region)]
provided.
(b) Upon doping, $|C_S(1)|$ is shown as a function of $\delta$ for $L=6$ system at $T\simeq0.24$
(the lowest temperature reachable by DQMC, before the sign
problem becomes prohibitive; cf. \Fig{Fig:sign}), 
with $D^\ast=800$-$1200$. Linear extrapolations $1/D^\ast\to0$ are performed, 
with the extrapolation values depicted as asterisk symbols. 
The detailed extrapolations at $\delta\simeq0.1, 0.3$ are shown in the inset.
In both panels, the DQMC results are also shown for comparison as depicted by the square symbols. 
}
\label{Fig:Cd}
\end{figure}

\section{B. Convergency check and extrapolation}
Here we provide detailed convergency check of the spin correlation functions shown in the main text.
In \Fig{Fig:Cd}(a), at half filled cases, we show the spin correlation function $C_S(d=1)$ for different system size $L=6,8$, 
versus temperature $T$ with various bond dimensions $D^*=400$-$900$. 
For better readability, $C_S$ is shifted upwards by 0.1, for $L=6$ system.
As shown, in the whole temperature regime, 
for both $L=6, 8$ systems, all $C_S$ curves lie on top of each other, 
showing good agreement with the DQMC data. 
In the inset, $C_S$ at a low temperature $T\simeq0.12$ are collected at various $D^\ast$, 
showing excellent convergency versus $1/D^\ast$. 
In \Fig{Fig:Cd}(b), $|C_S(d=1)|$ is shown as a function of hole doping $\delta$ for $L=6$ system, 
with $D^\ast=800$-$1200$, at $T\simeq0.24$. 
At each doping rate, the XTRG data exhibit good linearity with $1/D^\ast$, enabling us to perform a linear 
extrapolation to $1/D^\ast=0$. 
As shown, the extrapolation value shows good qualitative agreement with the DQMC results. 
In the inset, the detailed extrapolations $1/D^\ast\to0$ 
at around $\delta\simeq0.1$ and $\delta\simeq0.3$ are provided.

\section{C. Charge Correlations in the Doped Fermi-Hubbard Model}
\subsection{1. Hole-hole and hole-doublon correlations}

In this section, we provide more results on charge correlations
\begin{equation}
C_{hl}(d) = \tfrac{1}{N_d}\sum_{|{i}-{j}|=d}\langle\hat h_{i} \hat l_{j} \rangle -\langle\hat h_{i} \rangle\langle\hat l_{j} \rangle,
\end{equation}
with $l\in \{h,d\}$ corresponding to $\hat{l}\in \{\hat{h},\hat{d} \}$,
where $\hat h_{i} \equiv |0\rangle\langle0|$
and $\hat{d}_i \equiv \left|\uparrow\downarrow\rangle \langle\uparrow\downarrow\right|$
are projectors into the empty and double occupied states,
respectively.
We consider a $6\times6$ system and set $\mu=1.5$ throughout.

\begin{figure}[tb!]
\includegraphics[width=1\linewidth]{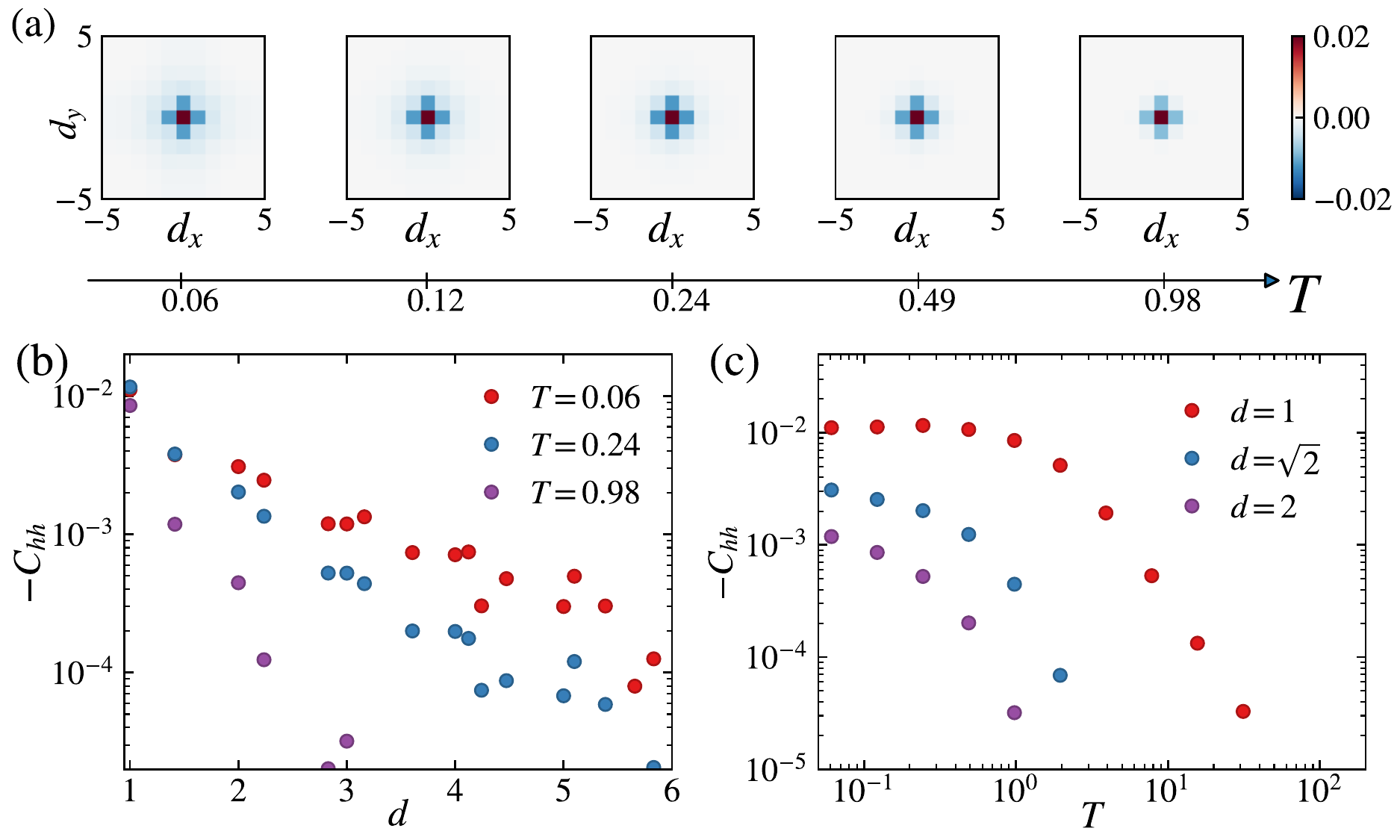}
\caption{\textit{Hole-hole correlation at doped cases.}
(a) Hole-hole correlation $C_{hh}(d)$, plotted as function 
of displacement $d_x, d_y$ along the horizontal and vertical directions, 
at various temperatures, for $6\times6$ system at fixed $\mu=1.5$ ($U=7.2$). 
(b) $C_{hh}(d)$ vs. $d$ at various temperatures,
and (c) $C_{hh}(d)$ vs. $T$ for various distances 
$d = 1, \sqrt{2}, 2$. 
}
\label{Fig:Chh}
\end{figure}

Figure \ref{Fig:Chh}(a) shows the hole-hole correlation
$C_{hh}$ plotted versus $d_x$ and $d_y$ 
from low (left) to high temperatures (right). 
There clearly exists a non-local anticorrelation 
in the spatial distribution, 
having $C_{hh}\le 0$ throughout,  
and decays roughly exponentially 
with distance for any fixed $T$
[\Fig{Fig:Chh}(b)]. 
When plotted vs. $T$ as in
\Fig{Fig:Chh}(c), 
the hole-hole anticorrelation persists to relatively high 
temperature [$T \lesssim 2$],
beyond which it rapidly
decays to zero. 
Note also that
around $T\simeq 2$ 
for given fixed $\mu=1.5$,
a maximal doping $\delta\simeq 0.17$ is reached
[see Fig.~\ref{Fig:Doped}(c) in the main text].
This appears naturally related to the
energy scale of the half-bandwidth $2t=2$
for the kinetic energy of the 2D square lattice
(ignoring the chemical potential since
$\delta\ll1$).

\begin{figure}[tb!]
\includegraphics[width=1\linewidth]{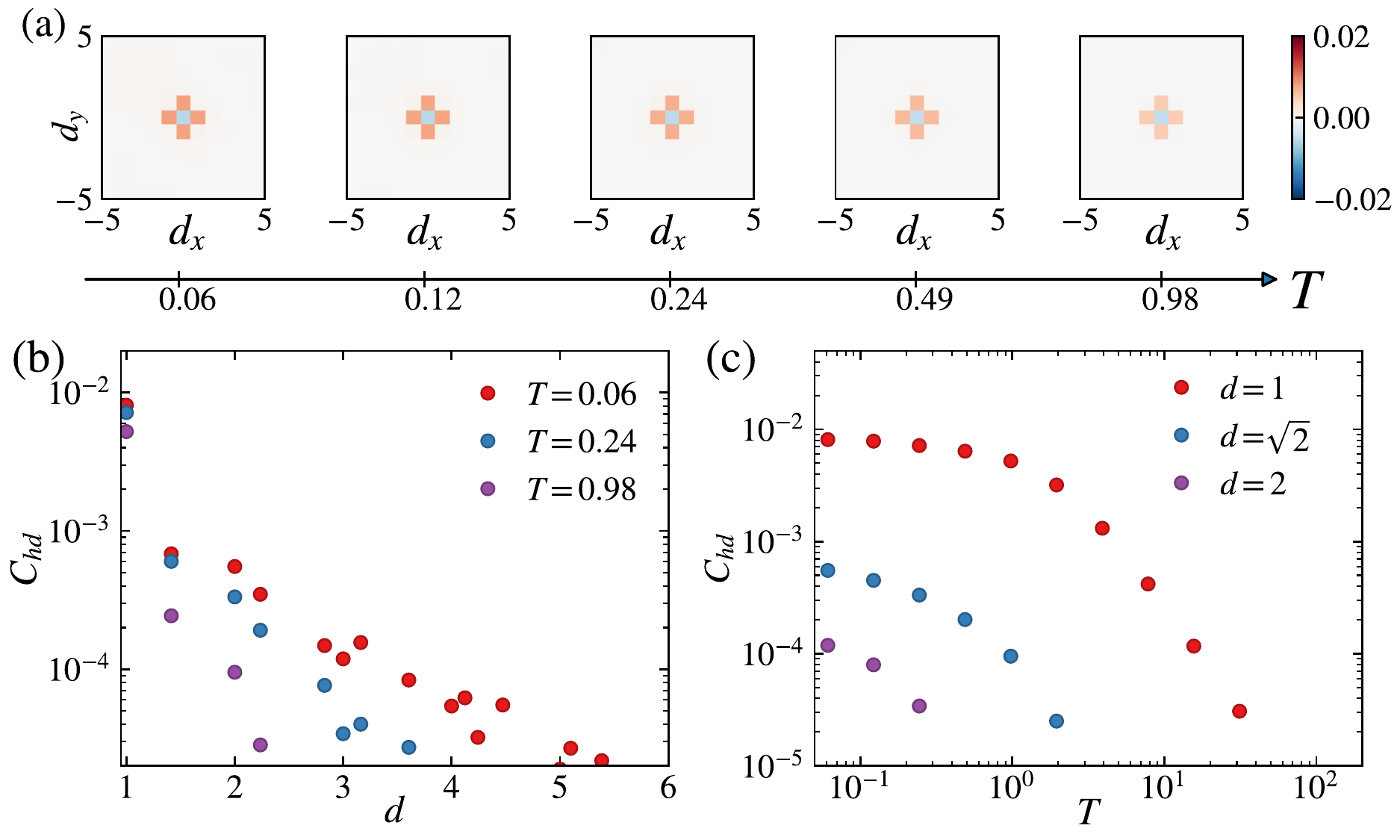}
\caption{\textit{Hole-doublon correlation at doped cases.}
Same layout as in \Fig{Fig:Chh} otherwise.
}
\label{Fig:Chd}
\end{figure}

\begin{figure}[tb]
\includegraphics[width=1\linewidth]{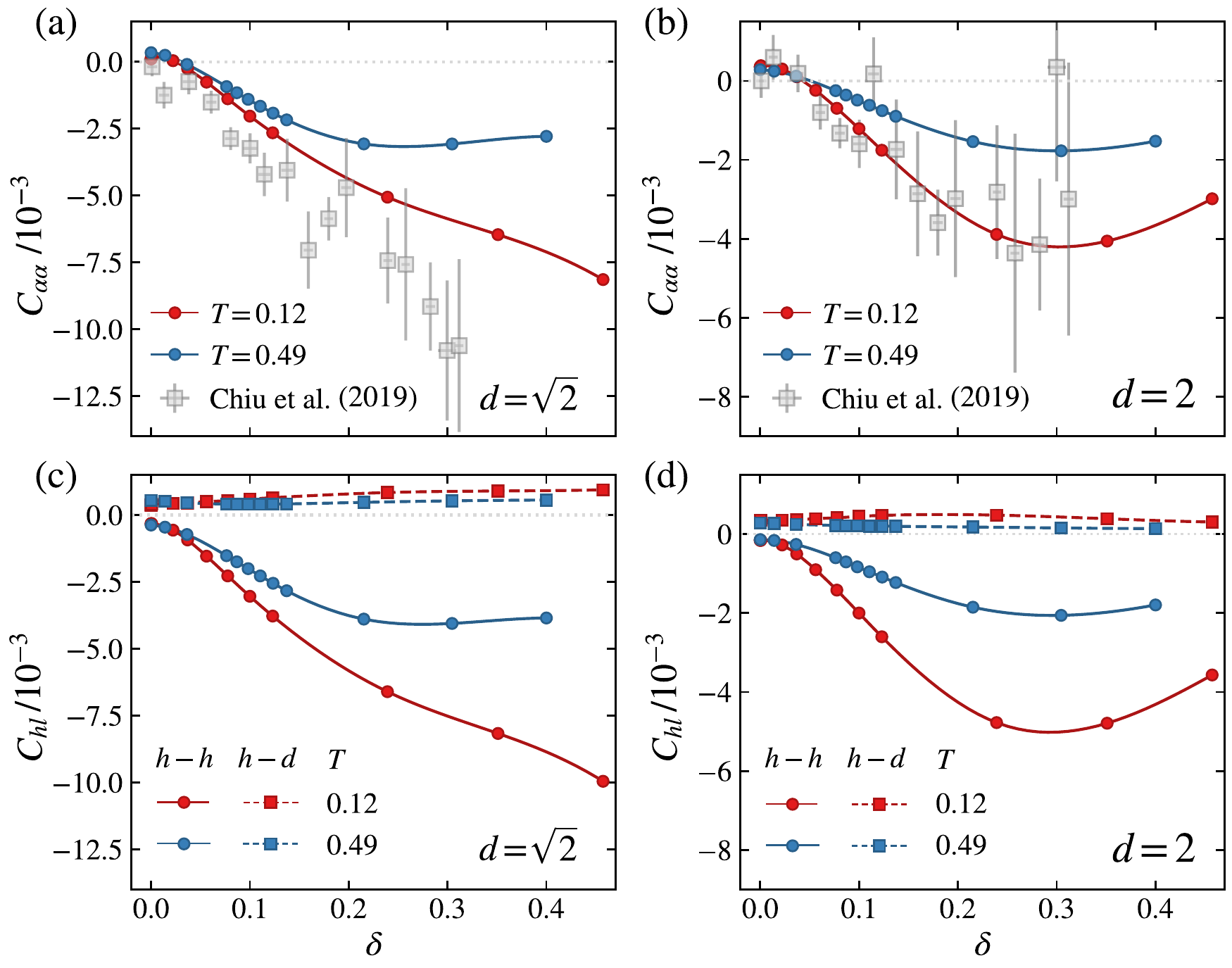}
\caption{\textit{Antimoment correlation functions.}
The antimoment correlations $C_{\alpha\alpha}(d)$
with (a) $d=\sqrt{2}$ and 
(b) $d=2$ are shown vs. $\delta$ at two different temperatures $T=0.12$ and $0.49$. 
The experimental data at $T/t\simeq0.25$ are also shown for comparison. 
The hole-hole (`h-h') and hole-doublon (`h-d') correlation function 
$C_{hl}$ with (c) $d=\sqrt{2}$ and (d) $d=2$ are shown versus doping $\delta$. 
The results are computed on a $6\times6$ square lattice ($U=7.2$).
}
\label{Fig:Ch}
\end{figure}

A completely analogous analysis is performed for
the hole-doublon correlations $C_{hd}$ 
as showns in \Fig{Fig:Chd}.
Figure \ref{Fig:Chd}(a) shows $C_{hd}$ vs. $d_x$ and $d_y$ at various temperatures,
where we observe nonlocal correlations between the hole-doublon pairs. 
Figure \ref{Fig:Chd}(b,c) shows that $C_{hd}$ decays rapidly 
with increasing distance $d$, and the hole-doublon correlation
again persist to a relatively high temperature
$T\sim 2$. 
Overall, the results in Figs.~\ref{Fig:Chh} and \ref{Fig:Chd} show that
the repulsive hole-hole and attractive hole-doublon pairs
are mainly limited to nearest-neighboring sites,
as expected given the sizable Coulomb interaction
$U=7.2$.

\subsection{2. Antimoment correlations}
In this section, we provide the results of antimoment correlation, 
\begin{subequations}\label{eq:Caa:def}
\begin{eqnarray}
   C_{\alpha\alpha}(d) 
 &=& \tfrac{1}{N_d} \sum_{|i-j|=d}  
   {\langle \hat \alpha_{i} \hat \alpha_{j}\rangle}-
   {\langle \hat \alpha_{i} \rangle\langle  \hat \alpha_{j} \rangle}
   \label{eq:Caa:def:1} \\
 &=& C_{hh} + 2C_{hd} + C_{dd}
\text{ ,}\label{eq:Caa:decomp}
\end{eqnarray}
\end{subequations}
with $\hat{\alpha}_{i} \equiv \hat h_{i} + \hat d_{i}$. This
can be directly compared with existing 
experimental  data \cite{Chiu2019} for 
$d=\sqrt{2}$ and $d=2$. 
Figure \ref{Fig:Ch}(a,b) shows $C_{\alpha\alpha}$ 
vs. doping $\delta$,
where a qualitative agreement with the experimental data can be observed. 
In both \Fig{Fig:Ch}(a,b), near half-filling 
a weak bunching effect is present. Thereafter the antimoments soon
exhibit strong anti-bunching effect as one dopes some holes into the system
($\delta\gtrsim3\%$ for $d=\sqrt{2}$ and $\delta\gtrsim5\%$ for $d=2$).

\begin{figure}[tp!]
\includegraphics[width=0.85\linewidth]{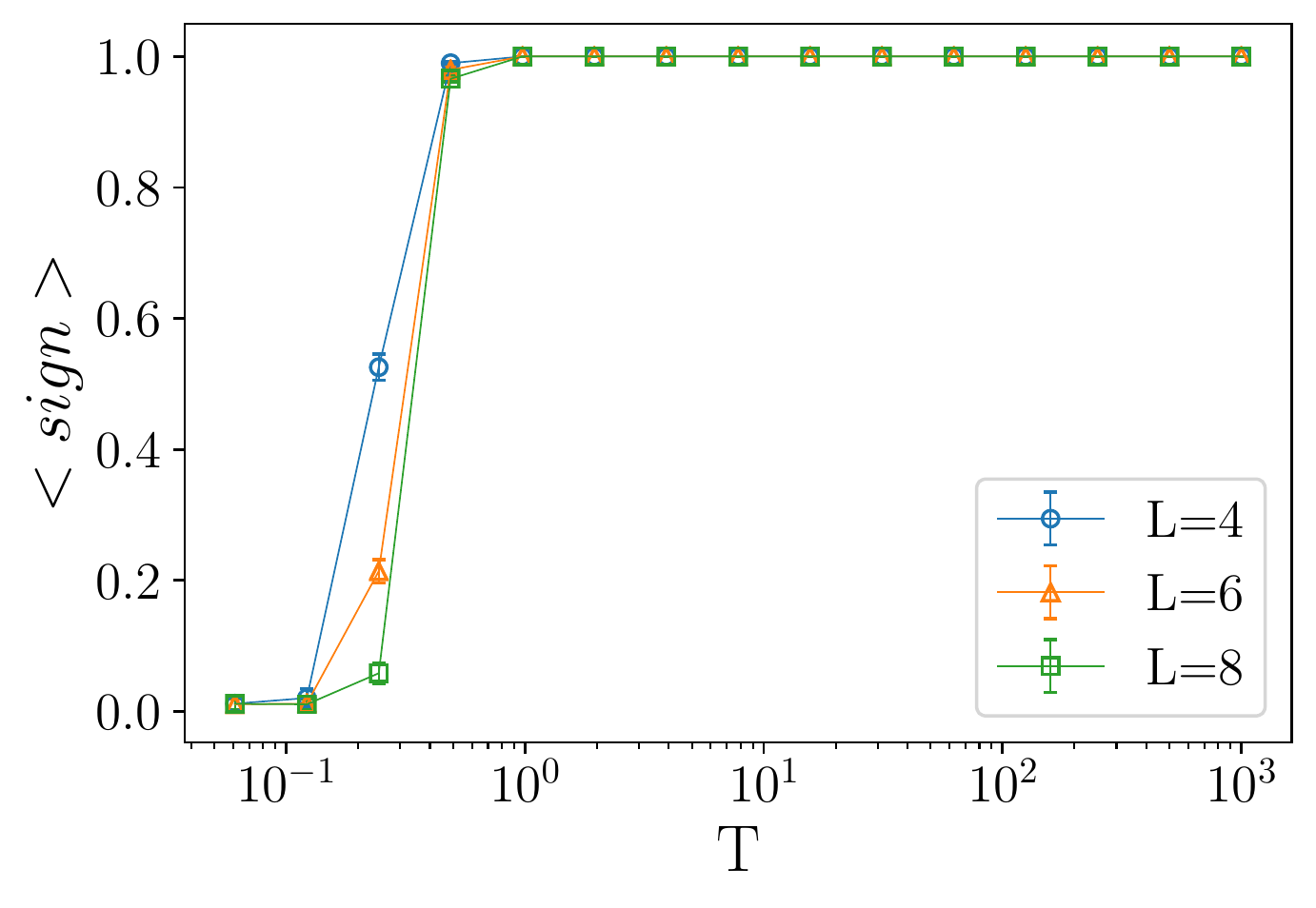}
\caption{\textit{DQMC average sign versus lattice size $L$ and temperature $T$.} 
In the calculations, the chemical potential is fixed at $\mu=1.5$, 
which corresponds to the data of Fig.~\ref{Fig:Doped} in the main text.}
\label{Fig:sign}
\end{figure}
 
Within XTRG, we can also compute all the partial contributions 
to the antimoment correlations as in \Eq{eq:Caa:decomp}.
The doublon-doublon correlation $C_{dd}$ is
negligibly small due to the rare density of doublons
considering hole-doping for large $U=7.2$. 
We thus only provide the results of $C_{hh}(d)$ and $C_{hd}(d)$ 
vs. doping in Fig.~\ref{Fig:Ch}(c,d).
Over the entire hole-doping 
regime considered in the present work,
the hole-hole correlation 
$C_{hh}(d)$ exhibits antibunching while the hole-doublon correlation $C_{hd}(d)$ 
exhibits bunching, for both $d=\sqrt{2}$ in Fig.~\ref{Fig:Ch}(c) and $d=2$ in Fig.~\ref{Fig:Ch}(d).
 
In the vicinity of half-filling, i.e., at low doping,
the hole-hole correlation $C_{hh}$
in \Fig{Fig:Ch}(c,d) becomes smaller (in absolute values)
as compared to the hole-doublon $C_{hd}>0$. This is responsible for the
bunching of antimoments at low doping [\Fig{Fig:Ch}(a,b)]. 
However, when more holes are doped into the system, e.g., 
$\delta \gtrsim 5$ \% as shown in \Fig{Fig:Ch}, 
the hole-hole repulsion becomes predominant and thus leads to the 
overall antibunching of antimoments.

\section{D. DQMC simulation and average sign in the doped cases}

We investigate the 2D square lattice Hubbard model with determinantal QMC simulations. The quartic term in Eq.~(\ref{Eq:Hub}) of the main text, 
$$U\hat{n}_{i\uparrow}\hat{n}_{i\downarrow}
=-\frac{U}{2}(\hat{n}_{i\uparrow}-\hat n_{i\downarrow})^2 +\frac{U}{2}(\hat{n}_{i\uparrow}+\hat{n}_{i\downarrow})
$$
is decoupled by a Hubbard-Stratonovich transformation to a form quadratic in $(\hat{n}_{i\uparrow} - \hat{n}_{i\downarrow})
= (\hat{c}^{\dagger}_{i\uparrow}\hat{c}_{i\uparrow}^{\ } - \hat{c}^{\dagger}_{i\downarrow}c_{i\downarrow}^{\ })$ coupled to an auxiliary Ising field on each lattice site~\cite{Blankenbecler1981}.
The particular decomposition above has the advantage
that the auxiliary fields can be chosen real.
The DQMC procedure obtains the partition function of the underlying Hamiltonian in a path integral formulation in a space of dimension $N = L \times L$ and an imaginary time $\tau$ up to $\beta = 1/T$. The auxiliary Ising field lives on the $L\times L\times \beta$ space-time configurational space and each specific configuration gives rise to one term in the configurational sum of the partition function. All of the physical observables are measured from the ensemble average over the space-time ($N\beta$) configurational weights of the auxiliary fields. As a consequence, the errors within the process are well controlled; specifically, the $(\Delta\tau)^2$ systematic error from the imaginary-time discretization, $\Delta\tau = \beta/M$, is controlled by the extrapolation $M \to\infty$ and the statistical error is controlled by the central-limit theorem.

The DQMC algorithm employed in this work is based on Ref.~\cite{Blankenbecler1981} and has been refined by including global moves to improve ergodicity and delay updating of the fermion Green function. This improves the efficiency of the Monte Carlo sampling~\cite{Assaad2008}. We have performed simulations for system sizes $L = 4, 6, 8$. The interaction is set as $U=7.2$ and we simulate temperatures from $T = 0.061$ to 1000 (inverse temperatures $\beta = 0.001$ to $16.39$).

We comment briefly on the sign problem in the Monte Carlo
sampling which becomes pronounced at finite doping. 
In general, the computational complexity in the presence of minus sign grows exponentially in the space-time volume  $N\beta$. This is because the correct physical observable now must include the effect of the sign of each Monte Carlo weight. One common practice is to use the absolute value of the weight to continue the Monte Carlo simulation, and then the expectation value becomes
$ \langle \hat{O} \rangle = \frac{\langle \hat{O} \cdot sign \rangle}{\langle sign \rangle}.$

Since the expectation value of the averaged sign, $\langle sign \rangle$, scales as $ e^{-\beta N}$, one cannot further extrapolate to the thermodynamic limit in this manner, as the error bar of any physical observables will explode. However, for finite size systems as investigated in this work, there is no problem of performing DQMC and obtaining unbiased results before $\langle sign \rangle$ becomes too small.  As shown in Fig.~\ref{Fig:sign}, for our system sizes $L=4,6,8$ at chemical potential of $\mu=1.5$ [cf. Fig.~\ref{Fig:Doped} (c) of the main text], the average sign is still affordable {down to} $T=0.244$ for $L=6,8$ [cf. \Fig{Fig:Cd}(b)].

Other DQMC parameters of the doped case, with minus sign problem in the main text, 
are investigated in a similar manner before the average sign becomes too small.

\end{document}